%% file: main.tex
\documentclass[10pt]{article}

\usepackage{amsmath}  
\usepackage{amssymb}  
\usepackage{array}  
\usepackage{authblk}  
\usepackage[backend=biber]{biblatex}
\usepackage{booktabs}  
\usepackage{colortbl}  
\usepackage{datetime}  
\usepackage{dsfont}   
\usepackage{float}  
\usepackage[colorlinks = true, linkcolor = blue, urlcolor = black, citecolor = blue]{hyperref}
\usepackage{mathtools}  
\usepackage{multirow}  
\usepackage{subfig}  
\usepackage{xcolor}
\usepackage{thmtools}
\usepackage{thm-restate}
\usepackage{graphicx}  
\usepackage{tabulary}  
\usepackage{booktabs}

\NewDocumentCommand{\NewSavedEnvironment}{m m}{
    \NewDocumentEnvironment{#1}{m +b}{%
        \begin{restatable}{donothing}{##1}
        \begin{#2}
        ##2
        \end{#2}
        \end{restatable}%
        }{\unskip\ignorespacesafterend}
    }
\NewSavedEnvironment{savedequation}{equation}

\newcommand{\Aone}{\hyperref[assumption:a1]{Assumption~(A1)}}
\newcommand{\Atwo}{\hyperref[assumption:a1]{Assumption~(A2)}}
\newcommand{\Athree}{\hyperref[assumption:a1]{Assumption~(A3)}}
\newcommand{\Afour}{\hyperref[assumption:a1]{Assumption~(A4)}}

\providecommand{\keywords}[1]{\textbf{\textit{Keywords:}} #1}

\graphicspath{{figures/}}

\bibliography{references.bib}

\title{%
    Resampling NANCOVA: Nonparametric \\
    Analysis of Covariance in Small Samples.%
}

\author[1,2,3]{\href{mailto:konstantin.thiel@pmu.ac.at}{Konstantin Emil Thiel}}
\author[4]{\href{mailto:paavo.sattler@tu-dortmund.de}{Paavo Sattler}}
\author[5]{\href{mailto:arne.bathke@plus.ac.at}{Arne C.\ Bathke}}
\author[1,2,3]{\href{mailto:georg.zimmermann@plus.ac.at}{Georg Zimmermann}}

\affil[1]{Research Programme Biomedical Data Science, Paracelsus Medical University, Salzburg, Austria}
\affil[2]{Faculty of Digital and Analytical Sciences, Paris Lodron University of Salzburg}
\affil[3]{Team Biostatistics and Big Medical Data, IDA Lab Salzburg, Paracelsus Medical University}
\affil[4]{Department of Statistics, TU Dortmund University, Germany}
\affil[5]{IDA Lab Salzburg, Department of Artificial Intelligence and Human Interfaces, Faculty of Digital and Analytical Sciences, Paris Lodron University of Salzburg}

\shortdate
\date{\today}

\begin{document}

\maketitle

\begin{abstract}
\noindent
Analysis of covariance is a crucial method for improving precision of statistical tests for factor effects in randomized experiments.
However, existing solutions suffer from one or more of the following limitations: (i) they are not suitable for ordinal data (as endpoints or explanatory variables); (ii) they require semiparametric model assumptions; (iii) they are inapplicable to small data scenarios due to often poor type-I error control; or (iv) they provide only approximate testing procedures and (asymptotically) exact test are missing.
In this paper, we investigate a resampling approach to the NANCOVA framework, which is a fully nonparametric model based on \emph{relative effects} that allows for an arbitrary number of covariates and groups, where both outcome variable (endpoint) and covariates can be metric or ordinal.
Thereby, we evaluate novel NANCOVA tests and a nonparametric competitor test without covariate adjustment in extensive simulations.
Unlike approximate tests in the NANCOVA framework, our resampling version showed good performance in small sample scenarios and maintained the nominal type-I error well.
Resampling NANCOVA also provided consistently high power: up to 26\% higher than the test without covariate adjustment in a small sample scenario with 4 groups and two covariates.
Moreover, we prove that resampling NANCOVA provides an asymptotically exact testing procedure, which makes it the first one in the NANCOVA framework.
In summary, resampling NANCOVA can be considered a viable tool for analysis of covariance that overcomes issues (i) - (iv).
\end{abstract}

\keywords{
    analysis of covariance, ANCOVA, covariate adjustment, baseline adjustment, nonparametric, rank-based, relative effect, resampling, bootstrap 
}

\input{sections/introduction}

\input{sections/preliminiaries}

\input{sections/resampling}

\input{sections/simulations}

\input{sections/data_example}

\input{sections/conclusion}

\section*{Acknowledgements}

GZ gratefully acknowledges the support of the WISS 2025 projects ``IDA-Lab Salzburg" (20204-WISS/225/197-2019 and 20102-F1901166-KZP) and ``EXDIGIT" (Excellence in Digital Sciences and Interdisciplinary Technologies) (20204-WISS/263/6-6022).
KET gratefully acknowledges the support of the WISS 2025 project ``ServEB" (20102/F2300645-FPR) and DEBRA Austria.

\appendix
\include{sections/proofs}

\pagebreak
\printbibliography

\end{document}

%% file: sections/introduction.tex
\section{Introduction}
\label{sec:introduction}

\emph{Covariate adjustment} is an essential technique to increase power of statistical tests for factor effects in randomized experiments. 
The key idea is to explain variability of the outcome variable using information from predictive covariates, which makes factor effects easier detectable.
In medical applications, covariate adjustment is considered such a relevant technique that there even exist guidelines from the European Medicines Agency~\cite{europeanmedicinesagencyGuidelineAdjustmentBaseline2015} and the US Food and Drug Administration~\cite{u.s.departmentofhealthandhumanservicesfoodanddrugadministrationAdjustingCovariatesRandomized2023}.

\begin{figure}[h]
  \centering
  \includegraphics[width=0.7\textwidth]{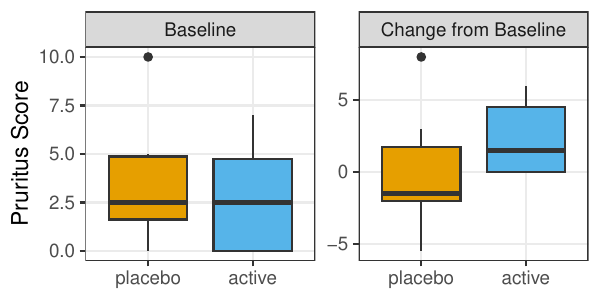}
  \caption{Pruritus score for $10$ patients under each treatment condition.}
  \label{fig:motivation}
\end{figure}

The conventional approach for covariate adjustment in randomized experiments is parametric \emph{analysis of covariance} (ANCOVA), which relies on mean values as estimands and presumes a linear model with normally distributed homoscedastic residuals.
Violation of these assumptions may compromise type-I error control and power of tests for factor effects~\cite{glassConsequencesFailureMeet1972, huitemaAnalysisCovarianceAlternatives2011}.
Even though there exist ANCOVA versions with relaxed distributional assumptions~\cite{sadooghi-alvandiParametricBootstrapApproach2013, zimmermannSmallsamplePerformanceUnderlying2019a}, these versions still rely on a linear model and on mean-based estimands, which requires metric data.
In many situations, however, data is measured on ordinal scales (e.g., \emph{pruritus scores} from a clinical trial shown in \autoref{fig:motivation}), and using mean values for ordinal variables is problematic.
Moreover, assuming a linear relationship between ordinal variables may be inappropriate, especially when the range of values is bounded or even discrete.

An appropriate estimand for ordinal data is the so-called nonparametric \emph{relative effect}~\cite{brunnerRankPseudoRankProcedures2018}, which is defined in the simple case of two groups as $q = P(X_1 \leq X_2) + 0.5 \cdot P(X_1 = X_2)$ for $X_1 \sim F_1$ and $X_2 \sim F_2$ independent. 
That is, $q$ measures the probability that observations from group 2 tend to be greater than observations from group 1.
Thas et al.~\cite{thasProbabilisticIndexModels2012} proposed \emph{probabilistic index models} (PIMs), where the relative effect is considered a function of covariates, and tests for factor effects can be implemented using dummy variables.
Still, determining the functional relationship in PIMs remains challenging.
Moreover, application in small sample studies is yet hampered by a less than satisfactory type I error control of PIMs.
While some progress has been made to improve the small sample performance~\cite{amorimSmallSampleInference2018, jaspersCovariateadjustedGeneralizedPairwise2024}, these solutions are only available for two-group comparisons.

One fully nonparametric approach designed for an arbitrary number of covariates and groups is Bathke and Brunner's \textbf{n}onparametric alternative to \textbf{an}alysis of \textbf{cova}riance~\cite{bathkeNonparametricAlternativeAnalysis2003}, which we refer to as NANCOVA.
NANCOVA corrects the relative effect in the outcome variable for random deviations of the relative effects in covariates, which leads to a \emph{covariate-adjusted relative effect} estimator with minimal variance.
However, only approximate tests for factor effects with liberal performance in small samples have been developed for NANCOVA until today.
The same holds for Schacht et al.'s $t$-approximation in a framework similar to NANCOVA, but with a different null hypothesis~\cite{schachtNewNonparametricApproach2008}.
The poor small sample performance of existing nonparametric approaches is particularly unsatisfactory as covariate adjustment would be crucial in these scenarios to improve the otherwise strictly limited power.

In this paper, we aim to overcome this issue by proposing a resampling approach to the NANCOVA framework of Bathke and Brunner.
Our resampling NANCOVA (i) controls type-I error well in small sample scenarios, (ii) provides a power gain compared to nonparametric tests without covariate adjustment, and (iii) provides the first asymptotically exact test in the NANCOVA framework.
We demonstrate (i) and (ii) in an extensive simulation study and address (iii) with a detailed proof.
Our approach performs resampling on the level of ranked observations, which allows for implementing bootstraps with a multitude of different \emph{random weights} that meet certain standard criteria~\cite{paulyWeightedResamplingMartingale2011}.
Unlike Friedrich et al.~\cite{friedrichWildBootstrapApproach2017}, who used Rademacher weights to implement a rank-based \emph{wild} bootstrap in a repeated measures scheme, we here propose to use multinomially distributed weights, which yields an \emph{Efron} bootstrap that exhibits both reliable type-I error control and competitive power in our simulations.

It is our intention to apply our resampling NANCOVA to the data example in \autoref{fig:motivation}, which is a simulated dataset based on a real randomized clinical trial that investigated a therapy for the rare skin disease \emph{epidermolysis bullosa}~\cite{wallyDiacereinOrphanDrug2018}.
Each patient's \emph{pruritus score} was recorded using a \emph{visual analogue scale} (VAS) at baseline, and the change from baseline was calculated after application of either a placebo cream or a cream that contained an active component. 
A recommended analysis option is to consider change from baseline as the outcome and use the baseline measurement as a covariate~\cite{europeanmedicinesagencyGuidelineAdjustmentBaseline2015}.
The combination of (i) the ordinal nature of both the raw pruritus scores and the pruritus change from baseline with (ii) the very small samples sizes make this dataset a perfect fit to demonstrate the potential advantages of our resampling NANCOVA.

\paragraph{Paper Outline.}

In the subsequent section we will recall basic definitions and introduce the NANCOVA model alongside with its technical assumptions.
In \autoref{sec:resampling-nancova}, we describe our bootstrap procedure and propose a corresponding test statistic.
Simulation results are presented in \autoref{sec:simulations}.
The example data from \autoref{fig:motivation} is analyzed using resampling NANCOVA in \autoref{sec:data-example}, where we also provide additional simulation results that include these real data.
Finally, we draw conclusions in \autoref{sec:conclusion}.
Moreover, proofs for all theoretical results are provided in \autoref{sec:proofs}.

%% file: sections/preliminiaries.tex
\section{Preliminaries}
\label{sec:preliminaries}

\subsection{Basic Definitions}

We consider independent $(d + 1)$-dimensional random vectors
\begin{equation}
  \label{eq:random-vectors}
  \mathbf{X}_{ik} = (X_{ik}^{(0)}, X_{ik}^{(1)}, \ldots, X_{ik}^{(d)})'
\end{equation}
where $i=1, \ldots a$ denotes the \emph{group} and $k = 1, \ldots, n_i$ denotes observations within group $i$.
The first component ($r = 0$) is the outcome variable, whereas components $r = 1, \dots, d$ are covariates.
It holds that ${X_{ik}^{(r)} \sim F_{i}^{(r)}}$, where $F_{i}^{(r)}$ is the \emph{normalized} cumulative distribution function (cdf) that is used to handle ties~\cite{munzelNonparametricMethodsMultivariate2000}.
With $N = \sum_{i = 1}^a n_i$ denoting the total sample size, we combine all samples in a matrix ${\mathbf{X} = (\mathbf{X}_{11}, \ldots, \mathbf{X}_{an_a})' \in \mathbb{R}^{N \times (d + 1)}}$.

\paragraph{Average Distribution and Rank Transforms.}

For each component, we define the average cdf ${H^{(r)} = \sum_{i=1}^a \nu_i F_i^{(r)}}$ as a convex combination of the group distributions.
For instance, $\nu_i = n_i / N$ yields a sample-size \emph{weighted} version and $\nu_i = 1 / a$ yields the \emph{unweighted} version.
Using the empirical cdf ${\widehat{F}_i^{(r)}}$~\cite{munzelNonparametricMethodsMultivariate2000}, we obtain the empirical counterpart ${\widehat{H}^{(r)} = \sum_{i=1}^a \nu_i \widehat{F}_i^{(r)}}$.
To construct a nonparametric effect measure, the average CDFs are applied to individual observations, which yields so-called \emph{asymptotic rank transforms} (ARTs) and \emph{rank transforms} (RTs):
\begin{align}
  \label{eq:rt}
  Y_{ik}^{(r)} = H^{(r)}(X_{ik}^{(r)})  &&%
  \widehat{Y}_{ik}^{(r)} = \widehat{H}^{(r)}(X_{ik}^{(r)})
\end{align}
While the ARTs $Y_{ik}^{(r)}$ are independent random variables, which is a useful property to derive theoretical results, they cannot be observed in practice as the average cdf $H^{(r)}$ is usually not known.
In order to construct estimators, the dependent RTs $\widehat{Y}_{ik}^{(r)}$ are used instead.
Note that the name rank transform is derived from the fact that ${\widehat{Y}_{ik}^{(r)} = \frac{1}{N}(R_{ik}^{(r)} - \frac{1}{2})}$, where $R_{ik}^{(r)}$ is the \emph{mid rank} of $X_{ik}^{(r)}$ when the sample-size weighted version of $\widehat{H}^{(r)}$ is used, whereas, in case of the unweighted version, $R_{ik}^{(r)}$ corresponds to the \emph{pseudo rank} of $X_{ik}^{(r)}$~\cite{brunnerRankPseudoRankProcedures2018}.

\paragraph{Relative Effects.}

The \emph{relative effect} defined as ${q_i^{(r)} = \int H^{(r)} d F_i^{(r)}}$ is the nonparametric effect measure used in NANCOVA and interpreted as follows:
If ${q_i^{(r)} < 0.5}$, then values from group $i$ tend to be smaller than the average, and if ${q_i^{(r)} > 0.5}$, then values from group $i$ tend to be greater than the average.
If ${q_i^{(r)} = 0.5}$, then there is no tendency towards smaller or larger values.
Consider the following expressions:
\begin{align}
  \label{eq:re-estimators}
  \bar{Y}_i^{(r)} = \frac{1}{n_i} \sum_{k = 1}^{n_i} Y_{ik}^{(r)}  &&%
  \widehat{q}_i^{\, (r)} = \frac{1}{n_i} \sum_{k = 1}^{n_i} \widehat{Y}_{ik}^{(r)} 
\end{align}
While $\widehat{q}_i^{\, (r)}$ is an unbiased and consistent estimator for $q_i^{(r)}$~\cite{brunnerRankPseudoRankProcedures2018}, its definition as a sum of dependent random variables makes the derivation of its asymptotic distribution cumbersome.
To this end, the \emph{asymptotically equivalent} expression $\bar{Y}_i^{(r)}$, which is defined as a sum of independent random variables, and therefore, allows for using standard central limit theorems, is used~\cite{munzelNonparametricMethodsMultivariate2000}.
For more details, we refer to \autoref{sec:resampling-nancova}.
To ease notation in the subsequent sections, we combine relative effects and mean ARTs in vectors
\begin{equation}
  \label{eq:art-vector}
  \bar{\mathbf{Y}} = ( \bar{\mathbf{Y}}'_1, \ldots, \bar{\mathbf{Y}}'_a )' = ( \bar{Y}_1^{(0)}, \ldots, \bar{Y}_1^{(d)}, \ldots, \bar{Y}_a^{(0)}, \ldots, \bar{Y}_a^{(d)} )'
\end{equation}
\begin{equation}
  \label{eq:re-vector}
  \widehat{\mathbf{q}} = ( \widehat{\mathbf{q}}'_1, \ldots, \widehat{\mathbf{q}}'_a )' = ( \widehat{q}_1^{\, (0)}, \ldots, \widehat{q}_1^{\, (d)}, \ldots, \widehat{q}_a^{\, (0)}, \ldots, \widehat{q}_a^{\, (d)} )'
\end{equation}
Analogously, we have ${\mathbf{q} = ( \mathbf{q}'_1, \ldots, \mathbf{q}'_a )' = \mathbb{E}(\widehat{\mathbf{q}}) = \mathbb{E}(\bar{\mathbf{Y}})}$.

\subsection{Technical Assumptions}
\label{sec:technical-assumptions}

NANCOVA neither requires a specific model nor a certain data distribution, yet there exist some assumptions that correspond to those in~\cite{bathkeNonparametricAlternativeAnalysis2003}:

\begin{itemize}
  \item[\textbf{(A1)}\label{assumption:a1}] $F_{i}^{(r)} = F_{j}^{(r)}$ for all $i, j = 1, \dots, a; r = 1, \dots, d$
  \item[\textbf{(A2)}\label{assumption:a2}] $\forall i = 1, \ldots, a : n_i/N \rightarrow \kappa_i \in (0, 1)$ for $N \rightarrow \infty$
  \item[\textbf{(A3)}\label{assumption:a3}] $\mathbf{S}_N \coloneqq \text{Cov}(\sqrt{N}\bar{\mathbf{Y}}) \rightarrow \mathbf{S}$ for $N \rightarrow \infty$
  \item[\textbf{(A4)}\label{assumption:a4}] $\exists \lambda_0 > 0 : \forall N \in \mathbb{N} : \lambda_{\text{min}}(\mathbf{S}_N) \geq \lambda_0$, where $\lambda_{\text{min}}(\mathbf{S}_N)$ is the smallest eigenvalue of $\mathbf{S}_N$.
\end{itemize}
\Aone{} says that marginal distributions are equal across groups, which should be fulfilled in a randomized trial, where subjects are randomly assigned to groups.
Note that \Aone{} implies that {$q_i^{(r)} = 1/2$} for all {$i = 1, \dots, a$} for each covariate {$r = 1, \dots, d$}.
\Atwo{} says that sample sizes in the individual groups grow at a similar rate, and eventually, the ratios $n_i/N$ converge.
Note that \Atwo{} is slightly more specific than the assumption $N/n_i \leq N_0 < \infty$ initially made in~\cite{bathkeNonparametricAlternativeAnalysis2003}.
However, since every bounded sequence comes with a convergent subsequence, we rather directly use the $\kappa_i$ notation in our assumptions.
\Athree{} states that the covariance matrix of the scaled ART vector converges, which is a standard assumption in nonparametric designs that was used by Bathke and Brunner to derive asymptotic normality of ${\sqrt{N}\bar{\mathbf{Y}}}$~\cite{bathkeNonparametricAlternativeAnalysis2003, munzelNonparametricMethodsMultivariate2000}.
Finally, \Afour{} states that the eigenvalues of $\mathbf{S}_N$ are bounded away from zero for all $N$, which guarantees that there is at least a minimal variance in all groups.

\subsection{Covariate Adjustment}
\label{sec:covariate-adjustment}

The idea of nonparametric covariate adjustment is to minimize the estimator's variance by adjusting $\widehat{q}_{i}^{\,(0)}$ for covariate effects in an arbitrary number of covariates.
We define the estimated covariate-adjusted relative effect of group $i = 1, \ldots, a$ as follows:
\begin{equation}
  \label{eq:cov-adj-re-general}
  \widehat{w}_i( \boldsymbol{\gamma} ) = \widehat{q}_{i}^{\,(0)} - \sum_{r = 1}^{d} \gamma^{(r)} \left ( \widehat{q}_{i}^{\,(r)} - \frac{1}{2} \right )
\end{equation}
Thereby, the coefficient vector ${\boldsymbol{\gamma} = (\gamma^{(1)}, \ldots, \gamma^{(d)})'}$ will later be chosen in a way that minimizes the variance of $\widehat{w}_i( \boldsymbol{\gamma} )$.
Definition \eqref{eq:cov-adj-re-general} can be interpreted as correcting $\widehat{q}_{i}^{\,(0)}$ for \emph{random deviations} of the covariate relative effects from their expectations.
Importantly, $\widehat{w}_i( \boldsymbol{\gamma} )$ combines both Bathke and Brunner~\cite{bathkeNonparametricAlternativeAnalysis2003}, who proposed adjustment for covariate effects instead of \emph{deviations} from their expectations, and Domhof~\cite{domhofNichtparametrischeRelativeEffekte2001}, who proposed this adjustment for deviations in a setting limited to $a = 2$ groups.

\paragraph{Variance Minimization.}

The coefficient vector $\boldsymbol{\gamma}$ is defined as the solution of the system of linear equations that minimizes the variance {$\text{Var}( \sum_{i = 1}^{a} \frac{n_i}{N} (1, -\boldsymbol{\gamma}')\bar{\mathbf{Y}}_i)$} of a weighted sum of adjusted ARTs~\cite{bathkeNonparametricAlternativeAnalysis2003}.
A consistent estimator $\widehat{\boldsymbol{\gamma}}$ is obtained using auxiliary terms
\begin{savedequation}{DefinitionEstimatorGammaC}
  \label{eq:definition-estimator-gamma-c}
  \widehat{C}^{(rs)} = \frac{1}{N} \sum_{i = 1}^a \sum_{k = 1}^{n_i} \left (\widehat{Y}_{ik}^{(r)} - \widehat{q}_i^{\, (r)} \right ) \left (\widehat{Y}_{ik}^{(s)} - \widehat{q}_i^{\, (s)} \right )
\end{savedequation}
for all $r, s = 0, \dots, d$. Then, the estimator is given by~\cite{bathkeNonparametricAlternativeAnalysis2003}
\begin{savedequation}{DefinitionEstimatorGammaMatrix}
  \label{eq:definition-estimator-gamma-matrix}
  \widehat{\boldsymbol{\gamma}} =%
  \begin{pmatrix}
    \widehat{C}^{(11)} & \widehat{C}^{(12)} & \cdots & \widehat{C}^{(1d)} \\
    \widehat{C}^{(21)} & \ddots & & \vdots \\
    \vdots & & & \\
    \widehat{C}^{(d1)} & \cdots & & \widehat{C}^{(dd)}
  \end{pmatrix}^{-1}%
  \begin{pmatrix}
    \widehat{C}^{(01)} \\
    \widehat{C}^{(02)} \\
    \vdots \\
    \widehat{C}^{(0d)}
  \end{pmatrix}
\end{savedequation}
To ease the notation, we will briefly write ${\widehat{\mathbf{w}} =  (\widehat{w}_1, \dots, \widehat{w}_a )' =  (\widehat{w}_1( \widehat{\boldsymbol{\gamma}} ), \dots, \widehat{w}_a( \widehat{\boldsymbol{\gamma}} )  )'}$ when we refer to the adjusted relative effects equipped with the specific coefficient estimator in \eqref{eq:definition-estimator-gamma-matrix}.
Note that we can define $\widehat{\mathbf{w}}$ also in matrix notation, which will further ease subsequent derivations.
To this end, we define an auxiliary matrix ${\widehat{\boldsymbol{\Gamma}} = \mathbf{I}_a \otimes ( 1, -\widehat{\boldsymbol{\gamma}}')}$ and an auxiliary vector ${\mathbf{e} = 1/2 \cdot \mathbf{1}_a \otimes (0, \mathbf{1}_d')'}$.
We may now write
\begin{equation}
  \label{eq:cov-adj-re-matrix-notation}
  \widehat{\mathbf{w}} = \widehat{\boldsymbol{\Gamma}} (\widehat{\mathbf{q}} - \mathbf{e})
\end{equation}

\subsection{Approximate NANCOVA}
\label{subsec:approximate-nancova}

Bathke and Brunner~\cite{bathkeNonparametricAlternativeAnalysis2003} proposed an ANOVA-type statistic (ATS), which is expected to perform better in small samples than Wald-type statistics (WTS)~\cite{brunnerBoxTypeApproximationsNonparametric1997}.
However, note that only approximate distribution properties can be derived for their ATS, that is, the proposed test is generally not an (asymptotically) exact one.

In nonparametric models, null hypotheses in tests for factor effects are usually formulated in terms of distribution functions~\cite{munzelNonparametricMethodsMultivariate2000, friedrichWildBootstrapApproach2017, brunnerRankPseudoRankProcedures2018}.
Hence, to facilitate inference about the outcome variable, we use the vector of outcome distributions ${\mathbf{F}^{(0)} = (F_{1}^{(0)}, \dots, F_{a}^{(0)})'}$.
With a suitable contrast matrix $\mathbf{K}$, that is, a matrix whose rows sum up to zero, the null hypothesis then reads
\begin{equation}
  \label{eq:h0-k}
  \mathcal{H}_{0}: \mathbf{K} \mathbf{F}^{(0)} = \mathbf{0}
\end{equation}
where $\mathbf{0}$ denotes the zero vector.
For example, the hypothesis of \emph{no factor effects}, that is, ${F_{1}^{(0)} = \ldots = F_{a}^{(0)}}$, can be formulated as ${\mathbf{K} \mathbf{F}^{(0)} = \mathbf{0}}$ with ${\mathbf{K} = \mathbf{I}_a - \frac{1}{a} \mathbf{J}_a}$, where $\mathbf{I}_a$ is the identity matrix and $\mathbf{J}_a$ is the ${a \times a}$ matrix with all entries equal to $1$.
The following properties indicate that that $\mathcal{H}_{0}$ can be tested using $\widehat{\mathbf{w}}$:
(i) \Aone{} implies ${\mathbb E (\widehat{\mathbf{w}}( \boldsymbol{\gamma} )) = \mathbf{q}^{(0)}}$, where ${\mathbf{q}^{(0)} = (q_{1}^{(0)}, \dots, q_{a}^{(0)})'}$ is the vector of relative effects in the outcome variable; (ii) it holds that ${\mathbf{K} \mathbf{q}^{(0)} \neq \mathbf{0}}$ implies ${\mathbf{K} \mathbf{F}^{(0)} \neq \mathbf{0}}$~\cite{brunnerRankPseudoRankProcedures2018}.

For the definition of the ATS, an idempotent and symmetric projection onto the column space of the contrast matrix $\mathbf{K}$ is required.
This projection is defined as ${\mathbf{T} = \mathbf{K}' (\mathbf{KK}')^{-} \mathbf{K}}$, where $(\mathbf{KK}')^{-}$ denotes a generalized inverse of $\mathbf{KK}$.
It holds that ${\mathbf{K} \mathbf{F}^{(0)} = \mathbf{0} \iff \mathbf{T} \mathbf{F}^{(0)} = \mathbf{0}}$~\cite{brunnerRankPseudoRankProcedures2018}.
The ATS is now defined as
\begin{savedequation}{AtsChiSquare}
  \label{eq:ats-chi-square}
  A_N = N \widehat{f} \cdot \frac%
  {\widehat{\mathbf{w}}' \mathbf{T} \widehat{\mathbf{w}}}%
  {\text{tr}(\mathbf{T} \widehat{\boldsymbol{\Sigma}})}
\end{savedequation}
where ${ \widehat{f} = {\text{tr}(\mathbf{T} \widehat{\boldsymbol{\Sigma}})^2} / {\text{tr}(\mathbf{T} \widehat{\boldsymbol{\Sigma}} \mathbf{T} \widehat{\boldsymbol{\Sigma}})} }$ is a degrees of freedom estimator and $\widehat{\boldsymbol{\Sigma}}$ is a consistent covariance estimator, which we define in \autoref{subsec:null-ditribution-an}.
$A_N$ is \emph{approximately} $\chi_{\widehat{f}}^2$-distributed under assumptions \hyperref[assumption:a1]{(A1)} -- \hyperref[assumption:a4]{(A4)}, and under ${\mathcal{H}_{0}: \mathbf{K} \mathbf{F}^{(0)} = \mathbf{0}}$~\cite{bathkeNonparametricAlternativeAnalysis2003}.

%% file: sections/resampling.tex
\section{Resampling NANCOVA}
\label{sec:resampling-nancova}

In this section, we will introduce our nonparametric resampling procedure and construct a bootstrap hypothesis test.
Our procedure is general and allows for implementation of an \emph{Efron bootstrap} as well as a \emph{wild bootstrap}~\cite{mammenBootstrapWildBootstrap1995}.
In this paper, however, we focus on the Efron bootstrap since it performed very well in our simulations.
We provide full proofs for all theorems that we subsequently state here in \autoref{sec:proofs}.

\paragraph{Motivation.}

Relative effect estimators $\widehat{\mathbf{q}}$ are sums of \emph{dependent} random variables, which makes standard central limit theorems inapplicable to derive the asymptotic distribution of $\sqrt{N} \widehat{\mathbf{q}}$.
In fact, one usually has to take cumbersome detours that involve the mean ARTs $\bar{\mathbf{Y}}$ as an unobservable auxiliary statistic.
It holds that $\widehat{\mathbf{q}}$  and $\bar{\mathbf{Y}}$ are \emph{asymptotically equal} in the sense that ${\sqrt{N} \mathbf{K} (\bar{\mathbf{Y}} - \widehat{\mathbf{q}}) \overset{P}{\rightarrow} 0}$ for a contrast matrix $\mathbf{K}$~\cite{munzelNonparametricMethodsMultivariate2000}.
Since each component of $\bar{\mathbf{Y}}$ is defined as sum of \emph{independent} random variables, an asymptotic normal distribution of ${\sqrt{N} \mathbf{K} \widehat{\mathbf{q}}}$ can eventually be derived.
The rationale of our bootstrap procedure is to implement a more direct way of approaching the asymptotic distribution using bootstrap samples.
Instead of defining a naive bootstrap counterpart to $\widehat{\mathbf{q}}$, we will define a bootstrap version of $\bar{\mathbf{Y}}$.
That is, we perform resampling on the level of rank transforms.
Note that Friedrich et al.\ implemented a similar approach in the context of repeated measures~\cite{friedrichWildBootstrapApproach2017}.

\paragraph{Bootstrapping Rank Transforms.}

We consider the set of rank transforms ${\widehat{\mathbf{Y}}_{11}, \ldots, \widehat{\mathbf{Y}}_{an_a}}$, where ${\widehat{\mathbf{Y}}_{ik} = (\widehat{Y}_{ik}^{(0)}, \ldots, \widehat{Y}_{ik}^{(d)})'}$ and $\widehat{Y}_{ik}^{(r)}$ is defined in~\eqref{eq:rt}, as the dataset from which we draw our bootstrap samples.
To implement an Efron bootstrap, $n_i$ samples are drawn with replacement from each group $i$.
We use an \emph{asterisk} (*) to denote resampled quantities, that is, we denote the $k$-th bootstrap sample drawn from group $i$ as
\begin{equation}
  \label{def:brt}
  \mathbf{Y}_{ik}^* = (Y_{ik}^{(0)*}, \ldots, Y_{ik}^{(d)*})'
\end{equation}
and refer to it as \emph{bootstrap rank transform (BRT)}.
Note that we could define the BRT also in terms of bootstrap samples on the level of the initial dataset.
Then, we'd have ${Y_{ik}^{(r)*} = \widehat{H}^{(r)}(X_{ik}^{(r)*})}$, where $X_{ik}^{(r)*}$ is the respective bootstrap sample.
From the bootstrap perspective, $\widehat{H}^{(r)}$ is the population distribution that does not change when bootstrap samples are re-drawn.
Hence, it holds that BRTs are conditionally \emph{independent}, that is, $\mathbf{Y}_{ik}^* \perp \mathbf{Y}_{j\ell}^*$ for $(i, k) \neq (j, \ell)$, which is analogous to ARTs defined in~\eqref{eq:rt}.
Indeed, BRTs can be considered a bootstrap counterpart to ARTs, which is the reason why we do not equip BRTs with a \emph{hat} as is the case for standard RTs defined in~\eqref{eq:rt}.
We now define \emph{mean BRTs}, which will serve as bootstrap counterparts to relative effect estimators as follows:
\begin{align}
  \label{eq:mean-brt}
  \begin{split}
    \bar{\mathbf{Y}}_i^* = \frac{1}{n_i} \sum_{k = 1}^{n_i} \mathbf{Y}_{ik}^* = \frac{1}{n_i} \sum_{k = 1}^{n_i} M_{ik} \cdot \widehat{\mathbf{Y}}_{ik}
  \end{split}
\end{align}
Thereby, ${(M_{i1}, \ldots, M_{in_i}) \sim \mathcal{M}(n_i, \frac{1}{n_i})}$ are \emph{multinomially distributed weights}, which correspond to the Efron bootstrap considered here.
To implement a wild bootstrap, instead, only the distribution of the weights ${M_{i1}, \ldots, M_{in_i}}$ must be adapted, for instance, to a \emph{Rademacher distribution}~\cite{mammenBootstrapWildBootstrap1995}.
Finally, to ease subsequent notation, we combine mean BRTs in a vector as follows:
\begin{equation}
  \label{eq:mean-brt-vector}
  \bar{\mathbf{Y}}^* = (\bar{\mathbf{Y}}_1^*, \ldots, \bar{\mathbf{Y}}_a^*)' = %
  (\bar{Y}_1^{(0)*}, \ldots, \bar{Y}_1^{(d)*}, \ldots \bar{Y}_a^{(0)*}, \ldots, \bar{Y}_a^{(d)*})'
\end{equation}
The following theorem forms the basis of our bootstrap procedure and shows that the conditional distribution of the scaled and centered BRT converges in probability to a normal distribution.
\begin{restatable}{theorem}{convergenceOfBRT}
  \label{theorem:convergence-of-brt}
  Under assumptions \Atwo{} and \Athree{}, conditionally on the sample $\mathbf{X}$, it holds that
  $$
  \sqrt{N} (\bar{\mathbf{Y}}^* - \widehat{\mathbf{q}}) \overset{d^*}{\longrightarrow} \mathcal{N}(\mathbf{0}, \mathbf{S}) \text{ in probability}
  $$
  where the covariance matrix $\mathbf{S}$ is given in \Atwo{}.
\end{restatable}

\noindent
Note that the distribution ${\mathcal{N}(\mathbf{0}, \mathbf{S})}$ is precisely the limiting distribution of the non-bootstrap counterpart ${\sqrt{N} (\bar{\mathbf{Y}} - \mathbf{q})}$.
Combined with the result on asymptotic equivalence ${\sqrt{N} \mathbf{K} (\bar{\mathbf{Y}} - \widehat{\mathbf{q}}) \overset{P}{\rightarrow} 0}$~\cite{munzelNonparametricMethodsMultivariate2000}, \autoref{theorem:convergence-of-brt} shows that our bootstrap statistic ${\sqrt{N} \mathbf{K} (\bar{\mathbf{Y}}^* - \widehat{\mathbf{q}})}$ mimics the unconditional null distribution of ${\sqrt{N} \mathbf{K} (\widehat{\mathbf{q}} - \mathbf{q})}$ under assumptions \hyperref[assumption:a1]{(A1)} -- \hyperref[assumption:a1]{(A4)}.

\paragraph{Estimators.}

In order to build a bootstrap test statistic that mimics the null distribution of the ATS $A_N$ defined in~\eqref{eq:ats-chi-square}, we need a bootstrap covariance estimator as well as additional bootstrap counterparts to the estimators $\widehat{\boldsymbol{\gamma}}$, and $\widehat{f}$.
Note that we omit the \emph{hat} for all bootstrap estimators to maintain similar notation for all BRT related quantities.
We define the bootstrap covariance estimator as follows:
\begin{savedequation}{DefinitionBsCovarianceEstimatorMatrix}
  \label{eq:definition-bs-covariance-estimator-matrix}
  \mathbf{S}^* = \bigoplus_{i = 1}^a \left [ \sigma_i^{(rs)*} \right ]^{r, s = 0, \ldots, d}
\end{savedequation}
where the {$(r, s)$-element} of the $i$-th diagonal block is defined as
\begin{savedequation}{DefinitionBsCovarianceEstimator}
  \label{eq:definition-bs-covariance-estimator}
  \sigma_i^{(rs)*} = \frac{N}{n_i(n_i - 1)} \sum_{k=1}^{n_i} (Y_{ik}^{(r)*} - \bar{Y}_{i}^{(r)*}) (Y_{ik}^{(s)*} - \bar{Y}_{i}^{(s)*})
\end{savedequation}
and ${\bar{Y}_{i}^{(r)*} = 1/n_i \sum_{i = 1}^{n_i} Y_{ik}^{(r)*}}$.
We obtain the bootstrap coefficient estimator by plugging-in BRTs into the respective formulas and obtain
\begin{equation}
  \label{eq:definition-bs-gamma-matrix}
  \boldsymbol{\gamma}^* = \begin{pmatrix}
    C^{(11)*} & C^{(12)*} & \cdots & C^{(1d)*} \\
    C^{(21)*} & \ddots &  & \vdots \\
    \vdots & & & \\
    C^{(d1)*} & \cdots & & C^{(dd)*}
  \end{pmatrix}^{-1}
  \begin{pmatrix}
    C^{(01)*} \\
    C^{(02)*} \\
    \vdots \\
    C^{(0d)*}
  \end{pmatrix}
\end{equation}
where
\begin{savedequation}{DefinitionBsGammaC}
  \label{eq:definition-bs-gamma-c}
  C^{(rs)*} = \frac{1}{N} \sum_{i = 1}^{a} \sum_{k=1}^{n_i} (Y_{ik}^{(r)*} - \bar{Y}_{i}^{(r)*}) (Y_{ik}^{(s)*} - \bar{Y}_{i}^{(s)*})
\end{savedequation}

\noindent
To simplify subsequent notation, we also define a matrix version of the coefficients
\begin{equation}
  \label{eq:def-auxiliary-bs-gamma-matrix}
  \boldsymbol{\Gamma}^* = \mathbf{I}_a \otimes \left ( 1, (-\boldsymbol{\gamma}^*)'\right )
\end{equation}
and an \emph{adjusted} version of the covariance matrix
\begin{equation}
  \label{eq:def-adjusted-bs-covariance-matrix}
  \boldsymbol{\Sigma}^* = \boldsymbol{\Gamma}^* \mathbf{S}^* (\boldsymbol{\Gamma}^*)'
\end{equation}
Finally, we obtain the bootstrap degrees of freedom estimator
\begin{equation}
  \label{eq:ats-bs-f-estimator}
  f^* = \frac%
  {\text{tr}(\mathbf{T} \boldsymbol{\Sigma}^*)^2}%
  {\text{tr}(\mathbf{T} \boldsymbol{\Sigma}^* \mathbf{T} \boldsymbol{\Sigma}^*)}
\end{equation}
The following theorem states that the bootstrap estimators are conditionally consistent:
\begin{restatable}{theorem}{conditionalConsistency}
  \label{theorem:conditional-consistency}
  Under assumptions \hyperref[assumption:a2]{(A2)} and \hyperref[assumption:a3]{(A3)}, conditionally on the sample $\mathbf{X}$, it holds that
  \begin{align*}
    \mathbf{S}^* &\xrightarrow{P^*} \mathbf{S} \text{ in probability} \\
    \boldsymbol{\gamma}^* &\xrightarrow{P^*} \boldsymbol{\gamma} \text{ in probability} \\
    f^* &\xrightarrow{P^*} f \text{ in probability}
  \end{align*}
\end{restatable}

\noindent
Thereby, the limiting covariance $\mathbf{S}$ is defined in \Athree{} and the limiting coefficients $\boldsymbol{\gamma}$ as well as the limiting degrees of freedom $f$ are defined in \autoref{sec:proofs}.

\paragraph{Resampling Test.}

Our bootstrap estimators allow for implementing the resampling test based on a bootstrap ATS.
Importantly, the bootstrap ATS must always approximate the null distribution of the non-bootstrap counterpart $A_N$. 
Therefore, we center ${\bar{\mathbf{Y}}^*}$ around its conditional mean $\widehat{\mathbf{q}}$, which leads to a centered \emph{bootstrap covariate-adjusted effect} $\boldsymbol{\Gamma}^* (\bar{\mathbf{Y}}^* - \widehat{\mathbf{q}})$. 
We can now define the bootstrap ATS as follows:
\begin{savedequation}{AtsSingleBootstrap}
  \label{eq:ats-single-bootstrap}
  A_N^* = N f^* \cdot \frac%
  {(\boldsymbol{\Gamma}^* (\bar{\mathbf{Y}}^* - \widehat{\mathbf{q}}))'\mathbf{T} \boldsymbol{\Gamma}^* (\bar{\mathbf{Y}}^* - \widehat{\mathbf{q}})}%
  {\text{tr} (\mathbf{T} \mathbf{\Sigma}^*)}
\end{savedequation}

\noindent
For inference, we compute the empirical distribution of $A_N^*$ using multiple bootstrap runs and use the $(1 - \alpha)$ quantile of this empirical distribution as the critical value for testing at the significance level $\alpha$.
The null hypothesis ${\mathcal{H}_0: \mathbf{K} \mathbf{F}^{(0)} = 0}$ is rejected if $A_N$ is greater than this critical value.
Importantly, this procedure yields an asymptotically exact test, which is the first one in the NANCOVA framework.
This property is formalized in the following theorem, which states that the conditional distribution of $A_N^*$ and the (unconditional) null distribution of $A_N$ converge to each other in the sense that the supremum norm between these distributions converges in probability to zero:
\begin{restatable}{theorem}{convergenceOfBootstrapATS}
  \label{theorem:convergence-of-bootstrap-ats}
  Under assumptions \hyperref[assumption:a1]{(A1)} -- \hyperref[assumption:a1]{(A4)}, the following holds:
  $$
  \sup_{x \in \mathbb R} \left | P(A_N^* \leq x | \mathbf{X} ) - P_{\mathcal{H}_0}(A_N \leq x) \right | \xrightarrow{P} 0
  $$
\end{restatable}

%% file: sections/simulations.tex
\section{Simulations}
\label{sec:simulations}

In this section, we empirically evaluate type-I error control and power of resampling NANCOVA using simulated data on both discrete ordinal and continuous metric scales.
Specifically, we focus on the small sample performance of an Efron Bootstrap version of resampling NANCOVA and compare it to existing approximate tests in the NANCOVA framework as well as to a nonparametric test without covariate adjustment.
The latter serves as a baseline to assess the power gain introduced by covariate adjustment.
Importantly, all power results will be discussed in the context of a test's ability to \emph{not exceed} the nominal type-I error rate.
All compared approaches test the null hypothesis ${\mathcal{H}_{0}: \mathbf{K} \mathbf{F}^{(0)} = \mathbf{0}}$.

\paragraph{Approximate Tests.}

In addition to the $\chi^2$ approximation from \autoref{subsec:approximate-nancova}, we investigate two additional ATS-based $\mathcal{F}$ approximations, one with and one without covariate adjustment.
Approximating the null distribution with an $\mathcal{F}$ distribution is recommended in small sample scenarios in nonparametric settings to improve type-I error control~\cite{brunnerBoxTypeApproximationsNonparametric1997}.
We explain the $\mathcal{F}$ approximation in the case with covariate adjustment (the test without covariate adjustment works analogously and its details can be found in~\cite{brunnerRankPseudoRankProcedures2018}):
Consider the test statistic ${\widetilde{A}_N = A_N / \widehat{f}}$, where $A_N$ and $\widehat{f}$ are defined in \eqref{eq:ats-chi-square}.
It holds that $\widetilde{A}_N$ can be approximated by a central ${\mathcal{F}(\widehat{f}, \widehat{f}_0)}$ distribution~\cite{brunnerBoxTypeApproximationsNonparametric1997, brunnerRankPseudoRankProcedures2018}.
Thereby, the degrees of freedom estimator is defined as
\begin{equation}
  \label{eq:ats-f0-estimator}
  \widehat{f_0} = \frac%
  {\text{tr} (\mathbf{D}_T \widehat{\mathbf{\Sigma}})^{2}}%
  {\text{tr} (\mathbf{D}_T^2 \widehat{\mathbf{\Sigma}}^2 \boldsymbol{\Lambda})}
\end{equation}
where $\mathbf{D}_T$ is the diagonal matrix of the diagonal elements of $\mathbf{T}$ and ${\boldsymbol{\Lambda} = \oplus_{i = 1}^{a} (n_i - 1)^{-1}}$.

\paragraph{Simulation Setup.}

We implemented and executed all simulations using the \texttt{R} programming language, version ${4.4.1}$~\cite{rcoreteamLanguageEnvironmentStatistical2024}.
Thereby, we implemented all tests in the NANCOVA framework from scratch, and we used the $\mathcal{F}$ approximation without covariate adjustment provided in the package \texttt{rankFD}~\cite{konietschkeRankFDRankBasedTests2022}.
Our simulations used sample-size weighted relative effects and we tested the null hypothesis of \emph{no group effect} by using the contrast matrix ${\mathbf{K} = \mathbf{I}_a - \frac{1}{a} \mathbf{J}_a}$.
The number of simulation runs was set to 5000 and our Efron bootstrap NANCOVA used 5000 bootstrap runs in each iteration.
We set the nominal $\alpha$ level to $5\%$ and use a $95\%$ Wald-confidence interval with this theoretical value of $\alpha$ and 5000 runs as a decision rule on whether the nominal type-I error rate is maintained.
This interval is ${[4.4, 5.6]}$.
For power simulations, we will discuss all results in the context of type-I error control in the sense of non-exceeding the upper bound of $5.6\%$.
To improve rendering of tables, we use the following abbreviations: (FA1) $\mathcal{F}$ approximation unadjusted (i.e., the test without covariate adjustment); (CA) $\chi^2$ approximation NANCOVA; (FA2) $\mathcal{F}$ approximation; (EB) Efron bootstrap NANCOVA.

\subsection{Discrete Ordinal Data}
\label{subsec:discrete-ordinal-data}

In this subsection, we provide simulation results on discrete ordinal data in a scenario with two groups and one covariate.
Simulation results with more groups and more covariates will be provided in \autoref{subsec:continuous-metric data}.
Discrete ordinal data frequently occurs in applied research, for instance, in data from questionnaires such as the \emph{Likert scale}.
In contrast to mean values, relative effects are interpretable also for ordinal data, which makes NANCOVA a natural choice as an analysis tool.

\paragraph{Data Generation.}

Ordinal data was generated using the \texttt{R} package \texttt{GenOrd}, version 1.4.0~\cite{barbieroGenOrdSimulationDiscrete2015}.
We let both the outcome variable $X^{(0)}$ and the covariate $X^{(1)}$ take $5$ different ordered values, say $v_1 < \ldots < v_5$.
Note that the precise numeric values are irrelevant to NANCOVA, as the values are replaced by ranks to estimate relative effects.
Association between $X^{(0)}$ and $X^{(1)}$ was specified by setting a correlation of $0.6$ in \texttt{GenOrd}'s \texttt{ordsample} function. 
For type-I error simulations, both $X^{(0)}$ and $X^{(1)}$ followed a discrete uniform distribution in both groups.
In power simulations, the covariate $X^{(1)}$ again followed a discrete uniform distribution in both groups, whereas the mariginal distribution of the outcome $X^{(0)}$ differed in both groups.
The exact outcome distributions in power simulations are specified in \autoref{tab:ordinal-outcome-distributions}, where it can be observed that the outcome im group 1 tended to smaller values and the outcome in group 2 tended to larger values.

\begin{table}[!h]
  \centering
  \caption{\label{tab:ordinal-outcome-distributions}Outcome distributions in the form of ${P( X^{(0)} = v_\ell )}$ for $\ell = 1, \ldots 5$ in power simulations with discrete ordinal data.}
  \begin{tabulary}{7cm}[t]{>{}l|CCCCC}
    Group & $v_1$ & $v_2$ & $v_3$ & $v_4$ & $v_5$ \\
    \midrule
    1 & 0.3 & 0.3 & 0.2 & 0.1 & 0.1 \\
    2 & 0.1 & 0.1 & 0.2 & 0.3 & 0.3 \\
    \midrule
  \end{tabulary}
\end{table}

\paragraph{Type-I Error.}

\autoref{tab:type-i-ordinal} displays the type-I error simulation results for 6 different sample size configurations.
The first three rows correspond to a total sample size of $N = 20$, while the last three rows correspond to a total sample size of $N = 40$.
Bathke and Brunner's $\chi^2$ approximation NANCOVA performed worst in type-I error control, as it exceeded the nominal rate in all scenarios with a worst case value of $10.46\%$ in the very small and highly unbalanced 5:15 scenario.
Even though $\mathcal{F}$ approximation NANCOVA was closer to the nominal rate than $\chi^2$ approximation NANCOVA, it still can't be considered a reliable test, as there was substantial deviation also in the balanced medium size 20:20 scenario.
However, it shall be noted that both $\chi^2$ approximation NANCOVA and $\mathcal{F}$ approximation NANCOVA appeared to converge to the nominal rate, since both were less liberal in medium size scenarios than in very small size scenarios.
$\mathcal{F}$ approximation unadjusted, on the other hand, came with rather good type-I error control already for very small samples.
Only in the very small and highly unbalanced 5:15 scenario, it was a too liberal test.
Finally, our Efron bootstrap NANCOVA never exceeded the nominal type-I error rate, and thus, can be best recommended for any application where false positive control is of high concern.
However, Efron bootstrap NANCOVA was somewhat conservative on very small sample sizes with a worst case deviation of $1.6\%$ below the nominal rate in the moderately unbalanced 8:12 scenario.
Still, Efron bootstrap NANCOVA seemed to converge to the nominal rate rather quickly, since it was already hardly conservative in the medium size scenarios.

\begin{table}[!h]
  \centering
  \caption{\label{tab:type-i-ordinal}Empirical type-I error on discrete ordinal data with $\alpha = 5\%$. Values exceeding a $95\%$ Wald interval are highlighted. Legend: (FA1) $\mathcal{F}$ approximation unadjusted; (CA) $\chi^2$ approximation NANCOVA; (FA2) $\mathcal{F}$ approximation NANCOVA; (EB) Efron bootstrap NANCOVA.}
  \begin{tabulary}{7cm}[t]{>{}l|CCCC}
    $n_1$:$n_2$ & FA1 & CA & FA2 & EB \\
    \midrule
    10:10 & 5.14 & \textcolor{magenta}{8.34} & \textcolor{magenta}{6.76} & 3.82\\
    8:12 & 4.70 & \textcolor{magenta}{8.16} & \textcolor{magenta}{5.98} & 3.40\\
    5:15 & \textcolor{magenta}{6.60} & \textcolor{magenta}{10.46} & \textcolor{magenta}{7.36} & 4.92\\
    \midrule
    20:20 & 4.68 & \textcolor{magenta}{6.14} & \textcolor{magenta}{6.42} & 4.64\\
    16:24 & 5.44 & \textcolor{magenta}{6.32} & 5.18 & 4.26\\
    10:30 & 5.58 & \textcolor{magenta}{7.66} & 5.36 & 4.92\\
    \midrule
  \end{tabulary}
\end{table}

\paragraph{Power.}

\autoref{tab:power-ordinal} shows power results in the same sample size configurations as in \autoref{tab:type-i-ordinal}.
A reasonable power comparison should not include tests that exceeded nominal the type-I error, as such tests could easily outperform competitors while never being practically applicable due to inflated false positive rate.
Hence, configurations where type-I error was exceeded are greyed out in \autoref{tab:power-ordinal}.
Consequently, in the 5:15 scenario, there was no competitor left for Efron bootstrap NANCOVA.
In the remaining very small sample size scenarios (10:10 and 8:12), Efron bootstrap NANCOVA outperformed $\mathcal{F}$ approximation unadjusted by roughly $10\%$ and in the medium sample size scenarios the power gain of Efron bootstrap NANCOVA over $\mathcal{F}$ approximation unadjusted was roughly $13$ - $15\%$.
In the two medium size scenarios where $\mathcal{F}$ approximation NANCOVA was applicable, it performed slightly better than Efron bootstrap NANCOVA, and thus, was the test with highest power.

\begin{table}[!h]
  \centering
  \caption{\label{tab:power-ordinal}Empirical power on discrete ordinal data with $\alpha = 5\%$. Configurations where the empirical type-I error substantially exceeds $\alpha$ are greyed out. Legend: (FA1) $\mathcal{F}$ approximation unadjusted; (CA) $\chi^2$ approximation NANCOVA; (FA2) $\mathcal{F}$ approximation NANCOVA; (EB) Efron bootstrap NANCOVA.}
  \begin{tabulary}{7cm}[t]{>{}l|CCCC}
    $n_1$:$n_2$ & FA1 & CA & FA2 & EB \\
    \midrule
    10:10 & 49.38 & \textcolor{lightgray}{73.30} & \textcolor{lightgray}{68.60} & 59.38\\
    8:12 & 46.68 & \textcolor{lightgray}{72.22} & \textcolor{lightgray}{64.68} & 56.26\\
    5:15 & \textcolor{lightgray}{40.98} & \textcolor{lightgray}{62.36} & \textcolor{lightgray}{51.64} & 40.82\\
    \midrule
    20:20 & 79.58 & \textcolor{lightgray}{94.84} & \textcolor{lightgray}{93.20} & 93.32\\
    16:24 & 77.64 & \textcolor{lightgray}{94.48} & 92.62 & 92.00\\
    10:30 & 63.66 & \textcolor{lightgray}{87.00} & 82.48 & 78.12\\
    \midrule
  \end{tabulary}
\end{table}

\subsection{Continuous Metric Data}
\label{subsec:continuous-metric data}

In this subsection, we present simulation results in scenarios with 4 groups and 2 covariates on continuous metric data.
We implemented such scenarios using a linear model with \emph{random} covariates and potential shift effects for power simulations.
While random covariates might be beyond the scope of standard theory for linear models~\cite{zimmermannSmallsamplePerformanceUnderlying2019a}, they are certainly covered by the NANCOVA framework.

\paragraph{Data Generation.}

Data was generated using the following linear model:
\begin{equation}
  X_{ik}^{(0)} = \mu_i + 2.5 (X_{ik}^{(1)} + X_{ik}^{(2)}) + \epsilon_{ik}
\end{equation}
That is, the outcome $X_{ik}^{(0)}$ was a linear combination of the covariates $X_{ik}^{(1)}$, $X_{ik}^{(2)}$, a group intercept $\mu_i$, and an independent error term $\epsilon_{ik}$.
$X_{ik}^{(1)}$ and $X_{ik}^{(2)}$ were independently uniformly distributed on the ${(0, 1)}$ interval and we allowed for three different distributions of $\epsilon_{ik}$: (i) a standard normal distribution; (ii) an exponential distribution with rate 1; and (iii) a heavily tailed $t$ distribution with 3 degrees of freedom.
In all cases (i) - (iii), we standardized $\epsilon_{ik}$ to zero mean and unit variance.
For type-I error simulations, we set $\mu_i = 0$ for all $i = 1, \ldots, 4$ and for power simulations, we set $\mu_3 = 0.5$ and $\mu_4 = 1$, while ${\mu_1 = \mu_2 = 0}$.

\paragraph{Type-I Error.}

\autoref{tab:lm-type-i} displays the empirical type-I error rates for the different tests on continuous metric data across 4 group sample configurations and our three error distributions (normal $\mathcal{N}$, exponential $\mathcal{E}$, and heavy-tailed t-distribution $t$).
The worst performance in terms of type-I error control was seen with the $\chi^2$ approximation NANCOVA, which exceeded the nominal rate in all scenarios, with the largest value of $9.50\%$ observed in the unbalanced 8:12:7:13 configuration under normally distributed errors.
Even in more balanced scenarios such as 10:10:10:10, $\chi^2$ approximation NANCOVA was too liberal, reaching a type-I error of $8.56\%$ under normal errors.
$\mathcal{F}$ approximation NANCOVA also struggled with controlling type-I error, particularly in smaller and unbalanced samples.
Its deviations were somewhat smaller than $\chi^2$ approximation NANCOVA, but still notable, such as $7.14\%$ in the balanced 10:10:10:10 configuration with normal errors and $6.22\%$ in the unbalanced 8:12:7:13 configuration with t-distributed errors.
However, it performed better in larger sample sizes, where it showed only mild deviations ($5.94\%$ in the 16:24:14:26 normal error scenario).
$\mathcal{F}$ approximation unadjusted generally demonstrated good type-I error control, with values close to the nominal rate across most configurations.
In balanced small sample size scenarios with non-normal errors, it became somewhat conservative, such as $3.90\%$ under t-distributed errors.
Efron bootstrap NANCOVA also provided good type-I error control, as it never exceeded the nominal rate in any configuration.
It was, however, slightly conservative in several settings, with the most conservative value being $3.22\%$ under exponential errors in the 10:10:10:10 configuration.

\begin{table}[!h]
  \centering
  \caption{\label{tab:lm-type-i}Empirical type-I error on continuous metric data with different error distributions and $\alpha = 5\%$. Values exceeding a $95\%$ Wald interval are highlighted. Legend: (FA1) $\mathcal{F}$ approximation unadjusted; (CA) $\chi^2$ approximation NANCOVA; (FA2) $\mathcal{F}$ approximation NANCOVA; (EB) Efron bootstrap NANCOVA.}
  \begin{tabulary}{8cm}[!h]{r|CCCC}
    $n_1$:$n_2$:$n_3$:$n_4$ & FA1 & CA & FA2 & EB \\
    \midrule
    \addlinespace[0.3em]
    10:10:10:10 & & & & \\
    \hspace{1em}$\mathcal{N}$ & 5.16 & \textcolor{magenta}{8.56} & \textcolor{magenta}{7.14} & 4.80\\
    \hspace{1em}$\mathcal{E}$ & 4.38 & \textcolor{magenta}{7.38} & \textcolor{magenta}{5.66} & 3.22\\
    \hspace{1em}$t$ & 3.90 & \textcolor{magenta}{8.20} & \textcolor{magenta}{5.90} & 4.28\\
    \addlinespace[0.3em]
    8:12:7:13 & & & & \\
    \hspace{1em}$\mathcal{N}$ & 4.96 & \textcolor{magenta}{9.50} & \textcolor{magenta}{6.10} & 4.76\\
    \hspace{1em}$\mathcal{E}$ & 4.84 & \textcolor{magenta}{9.12} & 5.32 & 3.96\\
    \hspace{1em}$t$ & 4.68 & \textcolor{magenta}{9.32} & \textcolor{magenta}{6.22} & 4.22\\
    \midrule
    \addlinespace[0.3em]
    20:20:20:20 & & & & \\
    \hspace{1em}$\mathcal{N}$ & 4.56 & \textcolor{magenta}{6.48} & 5.50 & 5.02\\
    \hspace{1em}$\mathcal{E}$ & 5.16 & \textcolor{magenta}{6.08} & 5.28 & 4.70\\
    \hspace{1em}$t$ & 5.10 & \textcolor{magenta}{6.14} & 5.48 & 4.94\\
    \addlinespace[0.3em]
    16:24:14:26 & & & & \\
    \hspace{1em}$\mathcal{N}$ & 5.58 & \textcolor{magenta}{6.80} & \textcolor{magenta}{5.94} & 5.06\\
    \hspace{1em}$\mathcal{E}$ & 4.78 & \textcolor{magenta}{6.06} & 4.82 & 4.30\\
    \hspace{1em}$t$ & 5.20 & \textcolor{magenta}{6.02} & 5.52 & 4.22\\
    \midrule
  \end{tabulary}
\end{table}

\paragraph{Power.}

\autoref{tab:lm-power} presents the power results for the different tests on continuous metric data under the same sample size configurations and error distributions as in \autoref{tab:lm-type-i}.
As before, tests that exceeded the nominal rate are greyed out since their power values should not be considered in a fair comparison.
In the balanced and unbalanced small sample configurations, Efron bootstrap NANCOVA showed a power advantage over $\mathcal{F}$ approximation unadjusted, outperforming it by roughly $15\%$ - $26\%$.
In medium-sized balanced samples (20:20:20:20), $\mathcal{F}$ approximation NANCOVA performed best across all error distributions.
However, Efron bootstrap NANCOVA performed competitive with power values reaching almost the level of $\mathcal{F}$ approximation NANCOVA.
This pattern persisted in the moderately unbalanced 16:24:14:26 scenario, where $\mathcal{F}$ approximation NANCOVA maintained a marginal power advantage over Efron bootstrap NANCOVA.
Compared to $\mathcal{F}$ approximation unadjusted, Efron bootstrap NANCOVA reached a power gain of roughly $31\%$ to $36\%$ in the balanced and unbalanced medium sample size scenarios.
Overall, Efron bootstrap NANCOVA remained a robust option with consistent high power.

\begin{table}[!h]
  \centering
  \caption{\label{tab:lm-power}Empirical power on continuous metric data with different error distributions and $\alpha = 5\%$. Configurations where the empirical type-I error substantially exceeds $\alpha$ are greyed out. Legend: (FA1) $\mathcal{F}$ approximation unadjusted; (CA) $\chi^2$ approximation NANCOVA; (FA2) $\mathcal{F}$ approximation NANCOVA; (EB) Efron bootstrap NANCOVA.}
  \begin{tabulary}{8cm}[!h]{r|CCCC}
    $n_1$:$n_2$:$n_3$:$n_4$ & FA1 & CA & FA2 & EB \\
    \midrule
    \addlinespace[0.3em]
    10:10:10:10 &&&& \\
    \hspace{1em}$\mathcal{N}$ & 25.66 & \textcolor{lightgray}{55.64} & \textcolor{lightgray}{48.96} & 43.22\\
    \hspace{1em}$\mathcal{E}$ & 28.18 & \textcolor{lightgray}{67.46} & \textcolor{lightgray}{61.06} & 54.66\\
    \hspace{1em}$t$ & 29.56 & \textcolor{lightgray}{70.54} & \textcolor{lightgray}{66.30} & 56.88\\
    \addlinespace[0.3em]
    8:12:7:13 &&&& \\
    \hspace{1em}$\mathcal{N}$ & 21.70 & \textcolor{lightgray}{53.62} & \textcolor{lightgray}{46.14} & 36.40\\
    \hspace{1em}$\mathcal{E}$ & 23.22 & \textcolor{lightgray}{62.36} & 54.76 & 45.22\\
    \hspace{1em}$t$ & 25.26 & \textcolor{lightgray}{68.74} & \textcolor{lightgray}{61.84} & 51.44\\
    \midrule
    \addlinespace[0.3em]
    20:20:20:20 &&&& \\
    \hspace{1em}$\mathcal{N}$ & 51.02 & \textcolor{lightgray}{85.06} & 83.46 & 82.24\\
    \hspace{1em}$\mathcal{E}$ & 58.06 & \textcolor{lightgray}{92.16} & 91.94 & 90.32\\
    \hspace{1em}$t$ & 59.74 & \textcolor{lightgray}{95.04} & 94.14 & 93.88\\
    \addlinespace[0.3em]
    16:24:14:26 &&&& \\
    \hspace{1em}$\mathcal{N}$ & 46.92 & \textcolor{lightgray}{83.42} & \textcolor{lightgray}{81.56} & 78.54\\
    \hspace{1em}$\mathcal{E}$ & 52.70 & \textcolor{lightgray}{91.72} & 89.28 & 88.10\\
    \hspace{1em}$t$ & 55.06 & \textcolor{lightgray}{94.46} & 93.02 & 91.40\\
    \midrule
  \end{tabulary}
\end{table}

%% file: sections/data_example.tex
\section{Data Example}
\label{sec:data-example}

In this section, we evaluate NANCOVA on the motivating data example mentioned in \autoref{sec:introduction}.
Thereby, \autoref{subsec:simulation} presents additional simulation results based on resampled versions of the dataset and \autoref{subsec:application} provides a data analysis from applying NANCOVA to the data example.

\paragraph{Dataset Description.}

\begin{figure}[!h]
  \centering
  \includegraphics[width=0.7\textwidth]{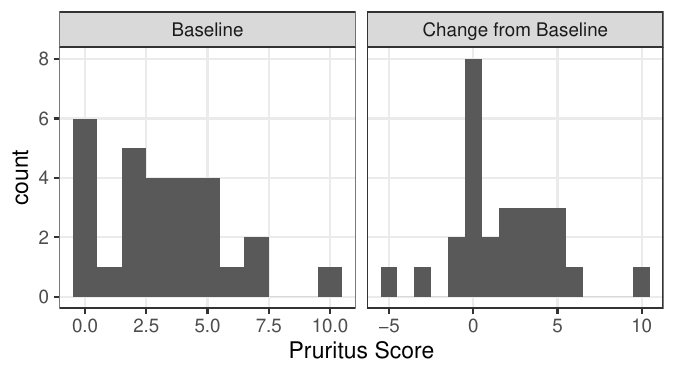}
  \caption{$N = 28$ pruritus measurements on a visual analogue scale according to Wally et al.~\cite{wallyDiacereinOrphanDrug2018}.}
  \label{fig:histogram}
\end{figure}

The data at hand was recorded in a randomized clinical trial on the rare skin disease epidermolysis bullosa by Wally et al.~\cite{wallyDiacereinOrphanDrug2018}.
Thereby, an active treatment was compared to a placebo in terms of \emph{pruritus scores}, which were measured using \emph{visual analogue scales} (VAS) at different points of time.
\autoref{fig:histogram} displays the raw data distribution, which exhibits zero inflation at both baseline and change from baseline.
Note that VAS scores are ordinal, and therefore, suitable for relative effect-based methods such as NANCOVA.
Precisely, we will use a simple model, where we consider change from baseline as the outcome and baseline as the covariate, which is a recommended approach in many contexts according to the European Medicines Agency~\cite{europeanmedicinesagencyGuidelineAdjustmentBaseline2015}.
In our case, the two variables are clearly associated with each other, which is confirmed by a sample correlation of $\widehat{\rho} = 0.68$, suggesting that covariate adjustment will actually help in explaining variability of the outcome variable.

\subsection{Simulation}
\label{subsec:simulation}

In this section, we evaluate NANCOVA's performance in terms of type-I error control and power in simulations based on the dataset from \autoref{fig:histogram}.
Thereby, we maintained the raw data characteristics as closely as possible by sampling outcome-covariate pairs from the full dataset and assigning these pairs to groups $i = 1, 2$.
In type-I error simulations, this created a situation, in which the null hypothesis ${\mathcal{H}_{0}: F_1^{(0)} = F_2^{(0)}}$, where $F_i^{(0)}$ is the change from baseline distribution in group $i = 1, 2$, was true.
For power simulations, we subtracted a Poisson distributed random effect with parameter $\lambda = 3$ from the change from baseline in group 2.
Thereby, we ensured that the VAS scores stayed in the ${[0, 10]}$ interval.
Note that a similar approach to simulating power on this dataset was recently implemented and supported by clinical expertise~\cite{geroldingerNeutralComparisonStatistical2024b}.
We used the same simulation parameters as in \autoref{sec:simulations} and the same sample size configurations as in \autoref{subsec:discrete-ordinal-data}.

\paragraph{Type-I Error.}

In terms of type-I error control, \autoref{tab:type-i-real-data} shows that both $\mathcal{F}$ approximation unadjusted and Efron bootstrap NANCOVA stood mostly within the nominal $\alpha = 5\%$ level, with values in the expected $95\%$ Wald interval.
Efron bootstrap NANCOVA was slightly conservative in some settings, notably showing a type-I error of $3.00\%$ in the 10:10 configuration and $2.90\%$ in the 8:12 configuration.
On the other hand, $\mathcal{F}$ approximation unadjusted tended to be consistent but again exceeded the nominal level in the unbalanced 5:15 scenario, where it reached $6.52\%$.
Notably, also Efron bootstrap NANCOVA mildly exceeded the Wald interval upper bound of $5.60\%$ in the unbalanced 10:30 scenario.
Both $\chi^2$ approximation NANCOVA and $\mathcal{F}$ approximation NANCOVA exhibited inflated type-I error rates across several configurations, with $\chi^2$ approximation NANCOVA showing the largest deviations, particularly in the 5:15 configuration, where it reached $12.60\%$.

\begin{table}[!h]
  \centering
  \caption{\label{tab:type-i-real-data}Empirical type-I error on real data with $\alpha = 5\%$. Values exceeding a $95\%$ Wald interval are highlighted. Legend: (FA1) $\mathcal{F}$ approximation unadjusted; (CA) $\chi^2$ approximation NANCOVA; (FA2) $\mathcal{F}$ approximation NANCOVA; (EB) Efron bootstrap NANCOVA.}
  \centering
  \begin{tabulary}{7cm}[t]{>{}l|CCCC}
    $n_1$:$n_2$ & FA1 & CA & FA2 & EB \\
    \midrule
    10:10 & 4.96 & \textcolor{magenta}{7.82} & 5.48 & 3.00\\
    8:12 & 5.28 & \textcolor{magenta}{9.14} & \textcolor{magenta}{6.28} & 2.90\\
    5:15 & \textcolor{magenta}{6.52} & \textcolor{magenta}{12.60} & \textcolor{magenta}{9.74} & 5.28\\
    \midrule
    20:20 & 5.10 & \textcolor{magenta}{6.66} & 5.50 & 4.40\\
    16:24 & 5.16 & \textcolor{magenta}{6.72} & \textcolor{magenta}{5.68} & 4.86\\
    10:30 & 5.20 & \textcolor{magenta}{8.16} & \textcolor{magenta}{7.70} & \textcolor{magenta}{5.72}\\
    \midrule
  \end{tabulary}
\end{table}

\paragraph{Power.}
In terms of power, \autoref{tab:power-real-data} shows that Efron bootstrap NANCOVA outperformed $\mathcal{F}$ approximation unadjusted in all scenarios where a comparison is possible, that is, configurations where both methods controlled the type-I error appropriately.
Precisely, the power gain of Efron bootstrap NANCOVA over $\mathcal{F}$ approximation unadjusted was approximately $7-9\%$, which was somewhat smaller than the power gain we observed in \autoref{sec:simulations}.
We account this to the fact that our simulation setup involved adding a randomly generated effect to the outcome in group 2, and thus, decreased the association between the covariate and the outcome, which caused the covariate to be less effective in explaining variability of the outcome.
In fact, while the average sample correlation $\widehat{\rho}$ between covariate and outcome was roughly $0.65$ - $0.66$ before adding the effect, it decreased to $0.55$ - $0.56$ after the effect was added.
Even though $\chi^2$ approximation NANCOVA and $\mathcal{F}$ approximation NANCOVA reported higher power in some cases, their inflated type-I error rates in several configurations make a fair comparison difficult.
In summary, Efron bootstrap NANCOVA demonstrated consistently better power than $\mathcal{F}$ approximation unadjusted while maintaining type-I error control, making it a more powerful alternative in all balanced and unbalanced designs where comparisons are valid.

\begin{table}[!h]
  \centering
  \caption{\label{tab:power-real-data}Empirical power on real data with $\alpha = 5\%$. Configurations where the empirical type-I error substantially exceeds $\alpha$ are greyed out. Legend: (FA1) $\mathcal{F}$ approximation unadjusted; (CA) $\chi^2$ approximation NANCOVA; (FA2) $\mathcal{F}$ approximation NANCOVA; (EB) Efron bootstrap NANCOVA.}
  \centering
  \begin{tabulary}{7cm}[t]{>{}l|CCCC}
    $n_1$:$n_2$ & FA1 & CA & FA2 & EB \\
    \midrule
    10:10 & 52.90 & \textcolor{lightgray}{75.96} & 71.38 & 61.16\\
    8:12 & 53.50 & \textcolor{lightgray}{73.80} & \textcolor{lightgray}{68.12} & 60.90\\
    5:15 & \textcolor{lightgray}{49.06} & \textcolor{lightgray}{67.06} & \textcolor{lightgray}{57.16} & 49.08\\
    \midrule
    20:20 & 84.90 & \textcolor{lightgray}{94.92} & 94.16 & 93.56\\
    16:24 & 83.90 & \textcolor{lightgray}{94.48} & \textcolor{lightgray}{93.92} & 92.14\\
    10:30 & 76.46 & \textcolor{lightgray}{89.06} & \textcolor{lightgray}{85.64} & \textcolor{lightgray}{79.40}\\
    \midrule
  \end{tabulary}
\end{table}

\subsection{Application}
\label{subsec:application}

\begin{figure}[!h]
  \centering
  \includegraphics[width=0.6\textwidth]{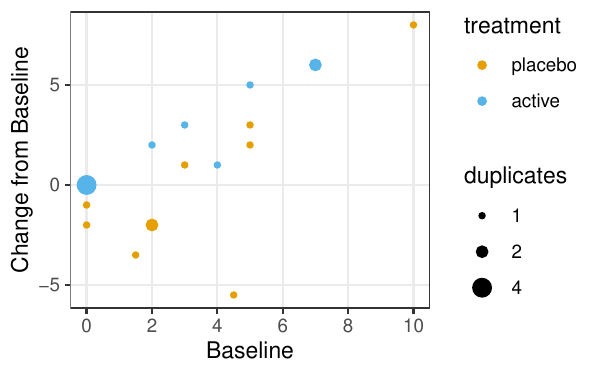}
  \caption{Scatterplot of the pruritus scores in \autoref{fig:motivation}.}
  \label{fig:scatterplot}
\end{figure}

We now want to apply the tests to the dataset from \autoref{fig:motivation}, which consists of $2 \times 10$ observations in 2 groups (placebo/active).
\autoref{fig:scatterplot} renders this dataset in a scatterplot and reveals some association between baseline and change from baseline.
In fact, we have a sample correlation of $\widehat{\rho} = 0.71$, which suggests high potential for covariate adjustment.
At first, however, we check the randomization assumption \hyperref[assumption:a1]{(A1)} by calculating relative effects of the covariate, that is, of the baseline measurement:
the estimated relative effects are $\widehat{q}_1^{\, (1)} = 0.52$ and $\widehat{q}_2^{\, (1)} = 0.48$, which is quite balanced and seems to confirm that randomization worked (in a perfectly balanced situation, we would observe ${\widehat{q}_1^{\, (1)} = \widehat{q}_2^{\, (1)} = 0.5}$).
We also calculate relative effects of the outcome, where we obtain $\widehat{q}_1^{\, (0)} = 0.38$ and $\widehat{q}_2^{\, (0)} = 0.62$.
Whether this tendency of change from baseline towards smaller values in the placebo group and greater values in the active group is significant will be answered by applying NANCOVA.

Therefore, we calculate the covariate coefficient according to \eqref{eq:definition-estimator-gamma-matrix}, which captures the rank correlation between outcome and baseline, and obtain $\widehat{\gamma}^{(1)} = 0.74$.
Using this, we calculate covariate-adjusted relative effects as follows:
\begin{align}
  \label{eq:cov-adjusted-rel-effect-real-data}
  \begin{split}
    \widehat{\mathbf{w}} = %
    \begin{pmatrix}
      0.38 \\
      0.62
    \end{pmatrix} - %
    0.74 \cdot %
    \begin{pmatrix}
      0.52 - 0.5 \\
      0.48 - 0.5
    \end{pmatrix} = \begin{pmatrix}
      0.37 \\
      0.63
    \end{pmatrix}
  \end{split}
\end{align}
As can be observed, $\widehat{\mathbf{w}}$ exhibits a slightly larger difference between the two groups than ${\widehat{\mathbf{q}}^{\, (0)} = (0.38, 0.62)'}$ does.
We interpret this as follows:
In a randomized setting, we have ${\mathbb{E} (\widehat{q}_1^{\, (1)}) = 0.5}$.
Since ${\widehat{q}_1^{\, (1)} = 0.52}$, the baseline values in the placebo group tend to be slightly larger than expected.
Recall from \autoref{fig:scatterplot} that baseline is positively correlated with change from baseline, which suggests that also the change from baseline values in the placebo group are slightly larger than expected.
That is, the relative effect $q_1^{\, (0)}$ should actually be slightly smaller on average than the estimate $\widehat{q}_1^{\, (0)} = 0.38$ suggests.
$\widehat{\mathbf{w}}$ corrects for this by capturing the correlation in the coefficient $\widehat{\gamma}^{(1)}$ and using it to decrease $\widehat{q}_1^{\, (0)}$ by the scaled deviation ${\widehat{\gamma}^{(1)} (\widehat{q}_1^{\, (1)} - \mathbb{E} (\widehat{q}_1^{\, (1)}))}$.
Analogously, this logic leads to the conclusion that change from baseline in the active group should be slightly larger on average than we observed here.
Hence, $\widehat{\mathbf{w}}$ increases $\widehat{q}_2^{\, (0)}$ by the scaled difference ${\widehat{\gamma}^{(1)} (\widehat{q}_2^{\, (1)} - \mathbb{E} (\widehat{q}_2^{\, (1)}))}$.

Finally, \autoref{tab:p-values} displays p-values of all tests that we previously simulated.
We observe that p-values of the tests with covariate-adjustment are consistently lower than the one of $\mathcal{F}$ approximation unadjusted, which conforms with our simulation results, where covariate adjustment increases the power.
In particular, Efron bootstrap NANCOVA allows for rejecting ${\mathcal{H}_{0}: F_1^{(0)} = F_2^{(0)}}$ at the ${\alpha = 5\%}$ level, which is in line with our simulation, whereas $\mathcal{F}$ approximation unadjusted is not able to reject.

\begin{table}[!h]
  \centering
  \caption{\label{tab:p-values}P-values from applying tests to the dataset from \autoref{fig:motivation}. Legend: (FA1) $\mathcal{F}$ approximation unadjusted; (CA) $\chi^2$ approximation NANCOVA; (FA2) $\mathcal{F}$ approximation NANCOVA; (EB) Efron bootstrap NANCOVA.}
  \centering
  \begin{tabulary}{7cm}[t]{CCCC}
    FA1 & CA & FA2 & EB \\
    \midrule
    0.075 & $< 0.001$ & $< 0.001$ & 0.013 \\
    \midrule
  \end{tabulary}
\end{table}

%% file: sections/conclusion.tex
\section{Conclusion}
\label{sec:conclusion}

In this paper, we have proposed resampling NANCOVA, a nonparametric method for covariate adjustment in randomized settings, applicable to metric as well as to ordinal data.
The capability for ordinal data originates from the fact that NANCOVA is based on relative effects, which measure the tendency towards larger or smaller values in the treatment groups.
Relative effects are estimated using ranks, and thus, can be computed for any data equipped with a suitable order relation.

\paragraph{Mathematical Foundation.}

Our resampling NANCOVA provides a general framework, which allows for implementing both an Efron bootstrap and a wild bootstrap.
We have proved that the resampling tests based on a scaled ANOVA-type statistic form an asymptotically exact testing procedure.
This includes conditional stochastic convergence of several involved quantities such as the bootstrapped covariance estimator.
To the best of our knowledge, our resampling test is the \emph{first} asymptotically exact testing procedure in the NANCOVA framework.
Existing NACOVA tests are approximative only, even for large samples~\cite{bathkeNonparametricAlternativeAnalysis2003, schachtNewNonparametricApproach2008}.

\paragraph{Rank-Based Efron Bootstrap.}

We have implemented and investigated a rank-based Efron bootstrap test in the NANCOVA framework, which forms an extension of existing resampling methods for relative effects~\cite{friedrichWildBootstrapApproach2017}.
In extensive simulations, our Efron bootstrap has yielded substantially better type-I error control than both an existing NANCOVA $\chi^2$ approximation and an additionally implemented $\mathcal{F}$ approximation.
Precisely, while the approximate tests have severely exceeded the nominal type-I rate, our Efron bootstrap has hardly ever exceeded the nominal rate, not even in very small and unbalanced sample size configurations.
Thereby, our Efron bootstrap test has yielded consistently better power than a relative effect-based test without covariate adjustment~\cite{konietschkeRankFDRankBasedTests2022}, which emphasizes the importance of performing covariate adjustment in nonparametric analyses.

\paragraph{Future Work.}

NANCOVA comes with the technical assumption of equal covariate distributions across treatment groups.
This assumption is expected to be met in randomized trials, but it requires a careful and thorough randomization procedure as otherwise errors might render NANCOVA hard to apply.
The reason is that NANCOVA works by adjusting the relative effect in the outcome variable for random deviations of the relative effects in the covariates from their means.
Importantly, these deviations must not exist due to systematic causes as the deviations might be transferred to the outcome and cause inflated type-I error.
In practice, it may be difficult to assess whether covariate distributions across treatment groups are balanced enough to allow the use of NANCOVA.
Therefore, we consider a mitigation of the randomization assumption an important direction for future research.

Currently, NANCOVA is designed for univariate outcomes only.
Investigating a NANCOVA version for multivariate outcomes by implementing a separate set of covariate coefficients for each outcome component might form another line of future research.

%% file: sections/proofs.tex
\section{Proofs}
\label{sec:proofs}

In this section, we provide proofs for the theorems stated in \autoref{sec:resampling-nancova}.
Additionally, in order to prove \autoref{theorem:convergence-of-bootstrap-ats}, we must derive the null distribution of the test statistic $A_N$ given in~\eqref{eq:ats-chi-square}.
Moreover, to complement preliminaries from \autoref{sec:preliminaries}, we define the covariance matrix used in~\eqref{eq:ats-chi-square} in the subsequent subsection.

\subsection{Covariance Estimator}
\label{subsec:covariance-estimator}

Before we introduce the covariance estimator, let's recall the definition of the ATS in \autoref{subsec:approximate-nancova}:

\AtsChiSquare*

\noindent
Thereby, ${ \widehat{f} = {\text{tr}(\mathbf{T} \widehat{\boldsymbol{\Sigma}})^2} ({\text{tr}(\mathbf{T} \widehat{\boldsymbol{\Sigma}} \mathbf{T} \widehat{\boldsymbol{\Sigma}})})^{-1} }$ is a degrees of freedom estimator.
The matrix ${\mathbf{T} \widehat{\boldsymbol{\Sigma}} \mathbf{T}}$ is used to estimate $\text{Cov}(\sqrt{N} \mathbf{T} \widehat{\mathbf{w}})$.
Recall from~\eqref{eq:cov-adj-re-matrix-notation} that ${\widehat{\mathbf{w}} = \widehat{\boldsymbol{\Gamma}} (\widehat{\mathbf{q}} - \mathbf{e})}$.
Thus, $\widehat{\boldsymbol{\Sigma}}$ is defined as
\begin{equation}
  \label{eq:derivation-covariance-estimator}
  \begin{split}
    \widehat{\boldsymbol{\Sigma}} &= \widehat{\boldsymbol{\Gamma}} \widehat{\mathbf{S}} \widehat{\boldsymbol{\Gamma}}'
  \end{split}
\end{equation}
where $\widehat{\boldsymbol{\Gamma}}$ is defined in~\eqref{eq:cov-adj-re-matrix-notation} and $\widehat{\mathbf{S}}$ is a block diagonal matrix consisting of ${a \times a}$ blocks of size ${(d + 1) \times (d + 1)}$, which is given by
\begin{savedequation}{DefinitionEmpiricalCovarianceMatrix}
  \label{eq:definition-empirical-covariance-matrix}
  \widehat{\mathbf{S}} = \bigoplus_{i = 1}^a \left [ \widehat{\sigma}_i^{(rs)} \right ]^{r, s = 0, \ldots, d}
\end{savedequation}
where
\begin{savedequation}{DefinitionEmpiricalCovariance}
  \label{eq:definition-empirical-covariance}
  \widehat{\sigma}_i^{(rs)} = \frac{N}{n_i(n_i - 1)} \sum_{k=1}^{n_i} \left ( \widehat{Y}_{ik}^{(r)} - \widehat{q}_{i}^{\, (r)} \right ) \left (\widehat{Y}_{ik}^{(s)} - \widehat{q}_{i}^{\, (s)} \right )
\end{savedequation}
It holds that ${\mathbf{T} \widehat{\boldsymbol{\Sigma}} \mathbf{T}}$ is consistent for the limiting covariance ${\lim_{N \rightarrow \infty} \text{Cov}(\sqrt{N} \mathbf{T} \widehat{\mathbf{w}})}$~\cite{munzelNonparametricMethodsMultivariate2000, bathkeNonparametricAlternativeAnalysis2003}.
Note that $\mathbf{T}$ being idempotent and symmetric implies ${\text{tr}(\mathbf{T} \widehat{\boldsymbol{\Sigma}}) = \text{tr}(\mathbf{T} \widehat{\boldsymbol{\Sigma}} \mathbf{T})}$ and ${\text{tr}(\mathbf{T} \widehat{\boldsymbol{\Sigma}} \mathbf{T} \widehat{\boldsymbol{\Sigma}}) = \text{tr}(\mathbf{T} \widehat{\boldsymbol{\Sigma}} \mathbf{T} \mathbf{T} \widehat{\boldsymbol{\Sigma}} \mathbf{T})}$ in~\eqref{eq:ats-chi-square}.

\subsection{Null Distribution of the ATS}
\label{subsec:null-ditribution-an}

To this date, the precise asymptotic null distribution of $A_N$ defined in~\eqref{eq:ats-chi-square} has not been derived.
In this subsection, we close this gap using standard results for relative effect and quadratic forms.
Firstly, we consider an arbitrary ${(c \times a)}$ contrast matrix, that is, a matrix for which ${\mathbf{K} \mathbf{1}_a = \mathbf{0}}$, and assume that the null hypothesis $\mathcal{H}_{0}: \mathbf{K} \mathbf{F}^{(0)} = \mathbf{0}$ holds.
Let ${\mathbf{T} = \mathbf{K}' (\mathbf{KK}')^{-} \mathbf{K}}$, where $(\mathbf{KK}')^{-}$ denotes a generalized inverse of $\mathbf{KK}$, be the idempotent and symmetric projection that is used in tests statistics (cf.\ \eqref{eq:ats-chi-square}).
Note that ${\mathbf{K} \mathbf{F}^{(0)} = \mathbf{0} \iff \mathbf{T} \mathbf{F}^{(0)} = \mathbf{0}}$.
Assuming \hyperref[assumption:a1]{(A1)} -- \hyperref[assumption:a1]{(A4)}, it can be concluded from Munzel and Brunner~\cite{munzelNonparametricMethodsMultivariate2000} that
\begin{equation}
  \label{eq:convergence-relative-effects}
  \sqrt{N} (\mathbf{T} \otimes \mathbf{I}_{d + 1}) (\widehat{\mathbf{q}} - \mathbf{q}) \overset{d}{\longrightarrow} \mathcal{N}(\mathbf{0}, (\mathbf{T} \otimes \mathbf{I}_{d + 1}) \mathbf{S} (\mathbf{T} \otimes \mathbf{I}_{d + 1}))
\end{equation}
where $\mathbb{E}(\widehat{\mathbf{q}}) = \mathbf{q}$ and $\mathbf{S}$ is the limiting covariance of the scaled ART vector given in~\eqref{eq:art-vector}, which is defined in \Athree{}.
Next, we define the limiting coefficient matrix
\begin{equation}
  \label{eq:def-auxiliary-limiting-gamma-matrix}
  \boldsymbol{\Gamma} = \mathbf{I}_a \otimes ( 1, -\boldsymbol{\gamma}' )
\end{equation}
where $\boldsymbol{\gamma}$ is the limit of $\widehat{\boldsymbol{\gamma}}$ defined in~\eqref{eq:definition-estimator-gamma-matrix}.
The following property holds:
\begin{align}
  \label{eq:quasi-commutativity-of-gamma-and-t}
  \begin{split}
    \mathbf{T} \boldsymbol{\Gamma} &= \mathbf{T} (\mathbf{I}_a \otimes ( 1, -\boldsymbol{\gamma}' )) \\
    &= \mathbf{T} \otimes ( 1, -\boldsymbol{\gamma}' ) \\
    &= (\mathbf{I}_a \otimes ( 1, -\boldsymbol{\gamma}' ))(\mathbf{T} \otimes \mathbf{I}_{d + 1}) \\
    &= \boldsymbol{\Gamma} (\mathbf{T} \otimes \mathbf{I}_{d + 1}) \\
  \end{split}
\end{align}
where we recall that $(\mathbf{T} \otimes \mathbf{I}_{d + 1})$ is the contrast matrix from~\eqref{eq:convergence-relative-effects}.
This property yields
\begin{align}
  \label{eq:convergence-numerator-with-limiting-gamma}
  \begin{split}
    \sqrt{N} \mathbf{T} \boldsymbol{\Gamma} (\widehat{\mathbf{q}} - \mathbf{q}) &= \sqrt{N} \boldsymbol{\Gamma} (\mathbf{T} \otimes \mathbf{I}_{d + 1}) (\widehat{\mathbf{q}} - \mathbf{q}) \\
    &\overset{d}{\longrightarrow} \mathcal{N} (\mathbf{0}, \boldsymbol{\Gamma} (\mathbf{T} \otimes \mathbf{I}_{d + 1}) \mathbf{S} (\boldsymbol{\Gamma} (\mathbf{T} \otimes \mathbf{I}_{d + 1}))') \\
    &= \mathcal{N}  (\mathbf{0}, \mathbf{T} \boldsymbol{\Gamma} \mathbf{S} \boldsymbol{\Gamma}' \mathbf{T}) \\
    &= \mathcal{N}  (\mathbf{0}, \mathbf{T} \boldsymbol{\Sigma} \mathbf{T})
  \end{split}
\end{align}
where we set ${\boldsymbol{\Sigma}} = \boldsymbol{\Gamma} \mathbf{S} \boldsymbol{\Gamma}'$.
Note that combining ${\mathbf{T} \mathbf{F}^{(0)} = \mathbf{0}}$ with \hyperref[assumption:a1]{(A1)} and result~\eqref{eq:quasi-commutativity-of-gamma-and-t} yields
\begin{equation}
  \label{eq:extended-contrast-property}
  \mathbf{T} \boldsymbol{\Gamma}\mathbf{q} = \mathbf{0}
\end{equation}
Next, we use results~\eqref{eq:convergence-numerator-with-limiting-gamma} and~\eqref{eq:extended-contrast-property} to derive the asymptotic distribution of a quadratic form similar to $A_N$.
Therefore, we also deploy (i) the fact that $\mathbf{T}$ is idempotent and symmetric, and (ii) standard results about the distribution of quadratic forms~\cite{mathaiQuadraticFormsRandom1992}, and obtain
\begin{align}
  \label{eq:convergence-quadratic-form-with-limiting-gamma}
  \begin{split}
    N f \cdot \frac%
    {(\boldsymbol{\Gamma} \widehat{\mathbf{q}})' \mathbf{T} \boldsymbol{\Gamma} \widehat{\mathbf{q}}}%
    {\text{tr}(\mathbf{T} \boldsymbol{\Sigma})} %
    &= f \cdot \frac%
    {(\sqrt{N} \mathbf{T} \boldsymbol{\Gamma} \widehat{\mathbf{q}})' \sqrt{N} \mathbf{T} \boldsymbol{\Gamma} \widehat{\mathbf{q}}}%
    {\text{tr}(\mathbf{T} \boldsymbol{\Sigma})} \\
    &= f \cdot \frac%
    {(\sqrt{N} \mathbf{T} \boldsymbol{\Gamma} (\widehat{\mathbf{q}} - \mathbf{q}))' \sqrt{N} \mathbf{T} \boldsymbol{\Gamma} (\widehat{\mathbf{q}} - \mathbf{q})}%
    {\text{tr}(\mathbf{T} \boldsymbol{\Sigma})} \\
    &\overset{d}{\longrightarrow} \frac{f}{\text{tr}(\mathbf{T \Sigma})} \sum_{i = 1}^a \lambda_i U_i
  \end{split}
\end{align}
where $\lambda_1, \ldots, \lambda_a$ are the eigenvalues of $\mathbf{T} \boldsymbol{\Sigma} \mathbf{T}$ and $U_1, \ldots, U_a \overset{\text{iid}}{\sim} \chi_1^2$.
Finally, expanding the ANOVA-type statistic $A_N$ defined in~\eqref{eq:ats-chi-square}, leveraging ${\mathbf{T} \widehat{\boldsymbol{\Gamma}} \mathbf{e} = \mathbf{0}}$, where $\mathbf{e}$ is the auxiliary vector used in~\eqref{eq:cov-adj-re-matrix-notation}, yields
\begin{align}
  \label{eq:a-n}
  \begin{split}
    A_N &= \widehat{f} \cdot \frac%
    {(\sqrt{N} \mathbf{T} \widehat{\mathbf{w}})' \sqrt{N} \mathbf{T} \widehat{\mathbf{w}}}%
    {\text{tr}(\mathbf{T} \widehat{\boldsymbol{\Sigma}})} \\
    &= \widehat{f} \cdot \frac%
    {(\sqrt{N} \mathbf{T} \widehat{\boldsymbol{\Gamma}} (\widehat{\mathbf{q}} - \mathbf{e}))' \sqrt{N} \mathbf{T} \widehat{\boldsymbol{\Gamma}} (\widehat{\mathbf{q}} - \mathbf{e})}%
    {\text{tr}(\mathbf{T} \widehat{\boldsymbol{\Sigma}})} \\
    &= \widehat{f} \cdot \frac%
    {(\sqrt{N} \mathbf{T} \widehat{\boldsymbol{\Gamma}} \widehat{\mathbf{q}})' \sqrt{N} \mathbf{T} \widehat{\boldsymbol{\Gamma}} \widehat{\mathbf{q}}}%
    {\text{tr}(\mathbf{T} \widehat{\boldsymbol{\Sigma}})} \\
    &= N \widehat{f} \cdot \frac%
    {(\widehat{\boldsymbol{\Gamma}} \widehat{\mathbf{q}})' \mathbf{T} \widehat{\boldsymbol{\Gamma}} \widehat{\mathbf{q}}}%
    {\text{tr}(\mathbf{T} \widehat{\boldsymbol{\Sigma}})} \\
    &\overset{d}{\longrightarrow} \frac{f}{\text{tr}(\mathbf{T \Sigma})} \sum_{i = 1}^a \lambda_i U_i
  \end{split}
\end{align}
where we combined the consistency of the estimators $\widehat{\boldsymbol{\Sigma}}$ from~\eqref{eq:derivation-covariance-estimator}, $\widehat{\boldsymbol{\Gamma}}$ 
from~\eqref{eq:cov-adj-re-matrix-notation}, and $\widehat{f}$ from~\eqref{eq:ats-chi-square} with convergence result~\eqref{eq:convergence-quadratic-form-with-limiting-gamma}, which concludes the derivation.

\subsection{Proof for \autoref{theorem:convergence-of-brt}}
\label{subsec:proof-theorem-convergence-of-brt}

In this subsection, we will provide our proof for the following theorem:
\convergenceOfBRT*

\noindent
Thereby, ${\bar{\mathbf{Y}}^*}$ denotes the mean bootstrap rank transform (BRT) vector defined in~\eqref{eq:mean-brt-vector} and ${\widehat{\mathbf{q}}}$ is the the relative effect vector defined in~\eqref{eq:re-vector}.
The proof will be conducted below, deploying a conditional central limit theorem.

\paragraph{Preliminaries.}

To execute the proof, we define auxiliary variables and start with \emph{centered} rank transforms (RTs) as follows:
\begin{equation}
  \label{eq:centred-rt}
  \widehat{Z}_{ik}^{(r)} = \widehat{Y}_{ik}^{(r)} - \widehat{q}_{i}^{\, (r)}
\end{equation}
Drawing bootstrap samples from the set of centered RTs yields \emph{centered} BRTs
\begin{equation}
  \label{eq:centred-brt}
  Z_{ik}^{(r)*} = Y_{ik}^{(r)*} - \widehat{q}_{i}^{\, (r)}
\end{equation}
where $\widehat{q}_{i}^{\, (r)}$ does not depend on the bootstrap sample, and hence, is not equipped with an asterisk.
To further ease notation, we will subsequently use vectors
\begin{equation}
  \label{eq:centred-rt-vector}
  \widehat{\mathbf{Z}}_{ik} = \left( \widehat{Z}_{ik}^{(0)}, \ldots, \widehat{Z}_{ik}^{(d)} \right)'
\end{equation}
and
\begin{equation}
  \label{eq:centred-brt-vector}
  \mathbf{Z}_{ik}^* = \left( Z_{ik}^{(0)^*}, \ldots, Z_{ik}^{(d)*} \right)'
\end{equation}

\paragraph{Proof.}

Note that ${\sqrt{N}(\bar{\mathbf{Y}}^* - \widehat{\mathbf{q}})}$ is a vector of $a$ blocks of size $(d + 1)$.
The $i$-th block can be re-written as
\begin{align}
  \label{eq:bootstrap-arts-written-out}
  \begin{split}
    \sqrt{N} (\bar{\mathbf{Y}}_i^* - \widehat{\mathbf{q}}_i)%
    &= \frac{\sqrt{N}}{n_i} \sum_{k = 1}^{n_i} \mathbf{Z}_{ik}^* \\
    &= \frac{\sqrt{N}}{n_i} \sum_{k = 1}^{n_i} M_{ik} \cdot \widehat{\mathbf{Z}}_{ik}
  \end{split}
\end{align}
where $\widehat{\mathbf{Z}}_{ik}$ is given in~\eqref{eq:centred-rt-vector} and $\mathbf{Z}_{ik}^*$ is given in~\eqref{eq:centred-brt-vector}, respectively, and $M_{ik}$ is an associated random weight.
In case of an Efron bootstrap, the weights are multinomially distributed, that is, ${(M_{i1}, \ldots, M_{in_i}) \sim \mathcal{M}\text{ultinomial}(n_i, \frac{1}{n_i})}$.

We now want to apply the conditional central limit theorem (CLT) derived in Pauly~\cite{paulyWeightedResamplingMartingale2011} to show, conditionally on $\mathbf{X}$, it holds that ${\sqrt{N} (\bar{\mathbf{Y}}_i^* - \widehat{\mathbf{q}}_i)}$ asymptotically follows a multivariate normal distribution.
Importantly, the conditional CLT works for any resampled statistic that can be written as a weighted sum of observations from the original sample, where the random weights meet the criteria stated in~\cite{paulyWeightedResamplingMartingale2011}.
Besides the multinomially distributed weights used in~\eqref{eq:bootstrap-arts-written-out}, this also applies, for instance, to Rademacher weights used to implement a \emph{wild} bootstrap.
Once suitable random weights have been selected, all further reasoning is only required at the level of the original sample.
Precisely, we have to check conditions (4.1) and (4.2) in~\cite{paulyWeightedResamplingMartingale2011}.
To verify (4.1), observe that the Euclidean norm of the scaled centered RT vector converges as follows:
\begin{align*}
  \left \lVert \frac{\sqrt{N}}{n_i} \widehat{\mathbf{Z}}_{ik} \right \rVert &= \frac{\sqrt{N}}{n_i} \sqrt{(\widehat{Z}_{ik}^{(0)})^2 + \ldots + (\widehat{Z}_{ik}^{(d)})^2} \\
  &\leq \frac{\sqrt{N}}{n_i} \sqrt{d + 1}  \tag*{since $\widehat{Y}_{ik}^{(r)} \leq 1$ for all $r = 0, \ldots, d$}\\
  &\xrightarrow{N \rightarrow \infty} 0  \tag*{by \Atwo{}}
\end{align*}
Hence, we have for $\varepsilon > 0$:
\begin{align*}
  P \left ( \max_{k \leq n_i} \, \left \lVert \frac{\sqrt{N}}{n_i} \widehat{\mathbf{Z}}_{ik} \right \rVert > \varepsilon \right ) &\leq %
  P \left ( \frac{\sqrt{N}}{n_i} \sqrt{d + 1} > \varepsilon \right ) %
  \xrightarrow{N \rightarrow \infty} 0
\end{align*}
In other words, ${\max_{k \leq n_i} \, \left \lVert \frac{\sqrt{N}}{n_i} \widehat{\mathbf{Z}}_{ik} \right \rVert \overset{P}{\longrightarrow} 0}$, and therefore, condition (4.1) in holds.
For condition (4.2), consider
\begin{align*}
  \frac{N}{n_i^2} \sum_{k=1}^{n_i} \widehat{\mathbf{Z}}_{ik} \cdot \widehat{\mathbf{Z}}_{ik}' %
  &= \left [ \frac{N}{n_i^2} \sum_{k=1}^{n_i} \widehat{Z}_{ik}^{(r)} \widehat{Z}_{ik}^{(s)} \right ]^{r, s = 0, \ldots, d} \\
  &= \left [ \frac{N}{n_i^2} \sum_{k=1}^{n_i} (\widehat{Y}_{ik}^{(r)} - \widehat{q}_{i}^{\, (r)})(\widehat{Y}_{ik}^{(s)} - \widehat{q}_{i}^{\, (s)}) \right ]^{r, s = 0, \ldots, d} \\
  &= \frac{n_i}{n_i - 1} \widehat{\mathbf{S}}_i
\end{align*}
where $\widehat{\mathbf{S}}_i$ denotes the $i$-th ${(d + 1) \times (d + 1)}$ diagonal block in the covariance estimator $\widehat{\mathbf{S}}$ defined in~\eqref{eq:definition-empirical-covariance-matrix}.
Munzel and Brunner showed the consistency result ${\widehat{\mathbf{S}} \xrightarrow{P} \mathbf{S}}$~\cite{munzelNonparametricMethodsMultivariate2000}, where $\mathbf{S}$ is defined in \Athree{}.
Thus, condition (4.2) in~\cite{paulyWeightedResamplingMartingale2011} is also met and we conclude based on the conditional CLT that, conditionally on $\mathbf{X}$,
\begin{equation*}
  \sqrt{N} (\bar{\mathbf{Y}}_i^* - \widehat{\mathbf{q}}_i) \overset{d^*}{\longrightarrow} \mathcal{N}(\mathbf{0}, \mathbf{S}_i) \hspace{3mm} \text{in probability}
\end{equation*}
where $\mathbf{S}_i$ is the $i$-th ${(d + 1) \times (d + 1)}$ diagonal block of the limiting covariance $\mathbf{S}$.
We recall that the bootstrap samples are drawn independently, and thus, conditionally on $\mathbf{X}$, it holds that ${\sqrt{N} (\bar{\mathbf{Y}}_i^* - \widehat{\mathbf{q}}_i) \perp \sqrt{N} (\bar{\mathbf{Y}}_j^* - \widehat{\mathbf{q}}_j)}$, for $i \neq j$.
Consequently, conditionally on $\mathbf{X}$, it also holds that $\sqrt{N}(\bar{\mathbf{Y}}^* - \widehat{\mathbf{q}}) \overset{d}{\longrightarrow} \mathcal{N}(\mathbf{0}, \mathbf{S})$ in probability, which concludes the proof of \autoref{theorem:convergence-of-brt}.

\subsection{Proof for \autoref{theorem:conditional-consistency}}
\label{subsec:proof-theorem-conditional-consistency}

In this subsection, we will provide our proof for the following theorem:
\conditionalConsistency*

\noindent
Thereby, the bootstrap estimators are defined in \autoref{sec:resampling-nancova}.
We start by proving conditional consistency of the covariance estimator $\mathbf{S}^*$, where we will deploy a conditional version of Chebyshev's inequality.
We will then be able to prove the other consistencies by tracing back to the consistency of the covariance estimator using continuity arguments.
Note that we conduct the proof using multinomially distributed weights that correspond to an Efron bootstrap (cf.~\autoref{sec:resampling-nancova}).

\subsubsection{Conditional Consistency of the Bootstrap Covariance Estimator}
\label{subsec:conditional-consistency-of-the-bootstrap-covariance-estimator}

Firstly, recall the definition of the bootstrap covariance estimator from \autoref{sec:resampling-nancova}:
\DefinitionBsCovarianceEstimatorMatrix*

\noindent
where the {$(r, s)$-element} of the $i$-th diagonal block is defined as
\DefinitionBsCovarianceEstimator*

\noindent
and $\bar{Y}_{i}^{(r)*}$ denotes the BRT mean.
On the other hand, the limiting covariance from \Athree{} is defined as
\begin{equation}
  \label{eq:definition-asymptotic-covariance-matrix}
  \begin{split}
  \mathbf{S} &= \lim_{N \rightarrow \infty} \bigoplus_{i=1}^a \text{Cov}(\sqrt{N} (\bar{\mathbf{Y}}_i - \mathbf{q}_i)) \\
  &\eqqcolon \bigoplus_{i = 1}^a \left [ \sigma_i^{(rs)} \right ]^{r, s = 0, \ldots, d}
  \end{split}
\end{equation}
where $\bar{\mathbf{Y}}_{i}$ is the $i$-th components of the ART vector from~\eqref{eq:art-vector} and $\mathbf{q}_i = \mathbb{E}(\bar{\mathbf{Y}}_{i})$.
The {$(r, s)$-element} of the $i$-th block is given by
\begin{equation}
  \label{eq:definition-asymptotic-covariance}
  \sigma_i^{(rs)} = \lim_{N \rightarrow \infty} \frac{N}{n_i} \text{Cov}(Y_{i1}^{(r)}, Y_{i1}^{(s)})
\end{equation}
In order to proof the conditional consistency of $\mathbf{S}^*$ in \autoref{theorem:conditional-consistency}, we will show that the following two statements hold:
\begin{align}
  \mathbb{E}(\sigma_i^{(rs)*} \, | \, \mathbf{X}) &\xrightarrow{P} \sigma_i^{(rs)}  \label{eq:convergence-expectaion-bs-covariance} \\
  \text{Var}(\sigma_i^{(rs)*} \, | \, \mathbf{X}) &\xrightarrow{P} 0  \label{eq:convergence-variance-bs-covariance}
\end{align}
From there, we will conclude the result by applying a conditional version of Chebyshev's inequality.

We start with the observation that
\begin{align}
  \label{eq:bs-covariance-convergence-equivalence-of-centering}
  \begin{split}
    &\sum_{k=1}^{n_i} (Y_{ik}^{(r)*} - \bar{Y}_{i}^{(r)*}) (Y_{ik}^{(s)*} - \bar{Y}_{i}^{(s)*}) \\
    &= \sum_{k=1}^{n_i} (Y_{ik}^{(r)*} - \widehat{q}_{i}^{\, (r)} - (\bar{Y}_{i}^{(r)*} - \widehat{q}_{i}^{\, (r)})) (Y_{ik}^{(s)*} - \widehat{q}_{i}^{\, (s)} - (\bar{Y}_{i}^{(s)*} - \widehat{q}_{i}^{\, (s)})) \\
    &= \sum_{k=1}^{n_i} (Z_{ik}^{(r)*} - \bar{Z}_{i}^{(r)*}) (Z_{ik}^{(s)*} - \bar{Z}_{i}^{(s)*})
  \end{split}
\end{align}
where $Z_{ik}^{(r)*}$ is the centered BRT defined in~\eqref{eq:centred-brt} and $\bar{Z}_{i}^{(r)*}$ is the respective mean.
Using \eqref{eq:bs-covariance-convergence-equivalence-of-centering}, we decompose \eqref{eq:definition-bs-covariance-estimator} as follows:
\begin{align}
  \sigma_i^{(rs)*} &= \frac{N}{n_i(n_i - 1)} \sum_{k=1}^{n_i} (Z_{ik}^{(r)*} - \bar{Z}_{i}^{(r)*}) (Z_{ik}^{(s)*} - \bar{Z}_{i}^{(s)*}) \notag \\
  &= \frac{N}{n_i(n_i - 1)} \sum_{k=1}^{n_i} \left ( %
  Z_{ik}^{(r)*} Z_{ik}^{(s)*} - \bar{Z}_{i}^{(s)*} Z_{ik}^{(r)*} - \bar{Z}_{i}^{(r)*} Z_{ik}^{(s)*} + \bar{Z}_{i}^{(r)*} \bar{Z}_{i}^{(s)*} %
  \right ) \notag  \\
  &= \frac{N}{n_i} \left ( \frac{1}{n_i - 1} \sum_{k=1}^{n_i} Z_{ik}^{(r)*} Z_{ik}^{(s)*} - \frac{n_i}{n_i - 1} \bar{Z}_{i}^{(r)*} \bar{Z}_{i}^{(s)*} \right ) \label{eq:bs-covariance-convergence-partial-result-0}
\end{align}
We will now investigate the expectation of the first term of \eqref{eq:bs-covariance-convergence-partial-result-0}.
Note that ${\sum_{k=1}^{n_i} Z_{ik}^{(r)*} Z_{ik}^{(s)*} = \sum_{k=1}^{n_i} M_{ik} \widehat{Z}_{ik}^{(r)} \widehat{Z}_{ik}^{(s)}}$, where $M_{ik}$ is the random weight introduced in~\eqref{eq:mean-brt} and $\widehat{Z}_{ik}^{(r)}$ is the centered RT from~\eqref{eq:centred-rt}.
Using the weight notation, we obtain
\begin{equation}
  \label{eq:bs-covariance-convergence-partial-result-1}
  \begin{split}
    \mathbb{E} \left ( \frac{N}{n_i(n_i - 1)} \sum_{k=1}^{n_i} Z_{ik}^{(r)*} Z_{ik}^{(s)*} \, | \, \mathbf{X} \right ) %
    &= \frac{N}{n_i(n_i - 1)} \sum_{k=1}^{n_i} \mathbb{E}(M_{ik}) \widehat{Z}_{ik}^{(r)} \widehat{Z}_{ik}^{(s)} \\
    &= \frac{N}{n_i(n_i - 1)} \sum_{k=1}^{n_i} (\widehat{Y}_{ik}^{(r)} - \widehat{q}_{i}^{\, (r)})(\widehat{Y}_{ik}^{(s)} - \widehat{q}_{i}^{\, (s)}) \\
    &= \widehat{\sigma}_i^{(rs)}
  \end{split}
\end{equation}
where $\widehat{\sigma}_i^{(rs)}$ is the covariance estimator defined in~\eqref{eq:definition-empirical-covariance}, which is consistent for $\sigma_i^{(rs)}$~\cite{munzelNonparametricMethodsMultivariate2000}.
Hence, to prove \eqref{eq:convergence-expectaion-bs-covariance}, it remains to show that the expectation of the second term in \eqref{eq:bs-covariance-convergence-partial-result-0} converges to zero.
Firstly, observe that the following proposition holds:
\begin{equation}
  \label{eq:bs-covariance-convergence-partial-result-2}
  \begin{split}
    \mathbb{E}(\bar{Z}_{i}^{(r)*} \, | \, \mathbf{X}) &= \frac{1}{n_i} \sum_{k=1}^{n_i} \mathbb{E} (M_{ik}) \widehat{Z}_{ik}^{(r)} \\
    &= \frac{1}{n_i} \sum_{k=1}^{n_i} \widehat{Y}_{ik}^{(r)} - \widehat{q}_{i}^{\, (r)} \\
    &= \widehat{q}_{i}^{\, (r)} - \widehat{q}_{i}^{\, (r)} = 0
  \end{split}
\end{equation}
Moreover, since $|\widehat{Z}_{ik}^{(r)}| \leq 1$, we have
\begin{equation}
  \label{eq:bs-covariance-convergence-partial-result-3}
  \begin{split}
    \mathbb{E}((\bar{Z}_{i}^{(r)*})^2 \, | \, \mathbf{X}) &\overset{\eqref{eq:bs-covariance-convergence-partial-result-2}}{=} \text{Var}(\bar{Z}_{i}^{(r)*} \, | \, \mathbf{X}) \\
    &= \text{Var} \left (\frac{1}{n_i} \sum_{k=1}^{n_i} M_{ik} \widehat{Z}_{ik}^{(r)} \, | \, \mathbf{X} \right ) \\
    &= \frac{1}{n_i^2} \left (\sum_{k=1}^{n_i} \text{Var}(M_{ik}) (\widehat{Z}_{ik}^{(r)})^2 + \sum_{\substack{k, \ell = 1; \\ k \neq \ell}}^{n_i} \text{Cov}(M_{ik}, M_{i \ell}) \widehat{Z}_{ik}^{(r)} \widehat{Z}_{i \ell}^{(r)} \right ) \\
    &\leq \frac{1}{n_i^2} \left (\sum_{k=1}^{n_i} (1 - \frac{1}{n_i}) (\widehat{Z}_{ik}^{(r)})^2 + \sum_{\substack{k, \ell = 1; \\ k \neq \ell}}^{n_i} \frac{1}{n_i} | \widehat{Z}_{ik}^{(r)} \widehat{Z}_{i \ell}^{(r)} | \right ) \\
    &\leq \frac{1}{n_i^2} \left ( n_i(1 - \frac{1}{n_i}) + n_i(n_i - 1) \frac{1}{n_i} \right ) \overset{N \rightarrow \infty}{\longrightarrow} 0
  \end{split}
\end{equation}
Applying Cauchy-Schwarz, and therein, leveraging \eqref{eq:bs-covariance-convergence-partial-result-3} yields
\begin{equation}
  \label{eq:bs-covariance-convergence-partial-result-4}
  \mathbb{E} ( \bar{Z}_{i}^{(r)*} \bar{Z}_{i}^{(s)*} \, | \, \mathbf{X} ) \overset{N \rightarrow \infty}{\longrightarrow} 0
\end{equation}
which concludes the proof of \eqref{eq:convergence-expectaion-bs-covariance}.

Next, we investigate the convergence of the conditional variance in \eqref{eq:convergence-variance-bs-covariance}.
Again, we investigate the terms of the decomposition \eqref{eq:bs-covariance-convergence-partial-result-0} separately and start with the first term.
\begin{equation}
  \label{eq:bs-covariance-convergence-partial-result-6}
  \begin{split}
    \text{Var} & \left ( \frac{1}{n_i - 1} \sum_{k=1}^{n_i} Z_{ik}^{(r)*} Z_{ik}^{(s)*} \, | \, \mathbf{X} \right ) %
    = \frac{1}{(n_i - 1)^2} \text{Var} \left ( \sum_{k=1}^{n_i} M_{ik} \widehat{Z}_{ik}^{(r)} \widehat{Z}_{ik}^{(s)} \, | \, \mathbf{X} \right ) \\
    &= \frac{1}{(n_i - 1)^2} \sum_{k=1}^{n_i} \text{Var}(M_{ik}) (\widehat{Z}_{ik}^{(r)} \widehat{Z}_{ik}^{(s)})^2 \\
    &+ \frac{1}{(n_i - 1)^2} \sum_{\substack{k, \ell = 1; \\ k \neq \ell}}^{n_i} \text{Cov}(M_{ik}, M_{i \ell}) \widehat{Z}_{ik}^{(r)} \widehat{Z}_{ik}^{(s)} \widehat{Z}_{i \ell}^{(r)} \widehat{Z}_{i \ell}^{(s)} \\
    &\leq \frac{1}{(n_i - 1)^2} \sum_{k=1}^{n_i} (1 - \frac{1}{n_i}) (\widehat{Z}_{ik}^{(r)} \widehat{Z}_{ik}^{(s)})^2 \\
    &+ \frac{1}{(n_i - 1)^2} \sum_{\substack{k, \ell = 1; \\ k \neq \ell}}^{n_i} \frac{1}{n_i} | \widehat{Z}_{ik}^{(r)} \widehat{Z}_{ik}^{(s)} \widehat{Z}_{i \ell}^{(r)} \widehat{Z}_{i \ell}^{(s)} | \\
    &\leq \frac{2}{(n_i - 1)} \overset{N \rightarrow \infty}{\longrightarrow} 0
  \end{split}
\end{equation}
For the second term, we leverage ${|\widehat{Z}_{ik}^{(r)}| \leq 1}$ to show
\begin{equation}
  \label{eq:bs-covariance-convergence-partial-result-7}
  \begin{split}
  \text{Var}( \bar{Z}_{i}^{(r)*} \bar{Z}_{i}^{(s)*} \, | \, \mathbf{X} ) &= %
  \mathbb{E}( (\bar{Z}_{i}^{(r)*} \bar{Z}_{i}^{(s)*} )^2 \, | \, \mathbf{X} ) - \mathbb{E}( \bar{Z}_{i}^{(r)*} \bar{Z}_{i}^{(s)*} \, | \, \mathbf{X} )^2 \\
  &\leq \mathbb{E}( (\bar{Z}_{i}^{(r)*} \bar{Z}_{i}^{(s)*} )^2 \, | \, \mathbf{X} ) \\
  &\leq \mathbb{E}( (1 \cdot \bar{Z}_{i}^{(s)*} )^2 \, | \, \mathbf{X} ) \\
  &\overset{N \rightarrow \infty}{\longrightarrow} 0
  \end{split}
\end{equation}
where the convergence follows from \eqref{eq:bs-covariance-convergence-partial-result-3}.
Now, with Cauchy-Schwarz' (C.S.) inequality in mind, consider
\begin{equation}
  \label{eq:bs-covariance-convergence-partial-result-8}
  \begin{split}
  \text{Var} & \left ( \frac{1}{n_i - 1} \sum_{k=1}^{n_i} Z_{ik}^{(r)*} Z_{ik}^{(s)*} - \frac{n_i}{n_i - 1} \bar{Z}_{i}^{(r)*} \bar{Z}_{i}^{(s)*} \, | \, \mathbf{X} \right ) \\
  &= \text{Var} \left ( \frac{1}{n_i - 1} \sum_{k=1}^{n_i} Z_{ik}^{(r)*} Z_{ik}^{(s)*} \, | \, \mathbf{X} \right ) \\
  &+ 2 \cdot \text{Cov} \left (\frac{1}{n_i - 1} \sum_{k=1}^{n_i} Z_{ik}^{(r)*} Z_{ik}^{(s)*} \, , \, %
  - \frac{n_i}{n_i - 1} \bar{Z}_{i}^{(r)*} \bar{Z}_{i}^{(s)*} \, | \, \mathbf{X} \right )  \\
  &+ \text{Var} \left (\frac{n_i}{n_i - 1} \bar{Z}_{i}^{(r)*} \bar{Z}_{i}^{(s)*} \, | \, \mathbf{X} \right )  \\
  &\overset{\text{C.S.}}{\leq} \text{Var} \left ( \frac{1}{n_i - 1} \sum_{k=1}^{n_i} Z_{ik}^{(r)*} Z_{ik}^{(s)*} \, | \, \mathbf{X} \right ) \\
  &+ 2 \cdot \sqrt{ \text{Var} \left ( \frac{1}{n_i - 1} \sum_{k=1}^{n_i} Z_{ik}^{(r)*} Z_{ik}^{(s)*} \, | \, \mathbf{X} \right ) \cdot %
  \text{Var} \left (\frac{n_i}{n_i - 1} \bar{Z}_{i}^{(r)*} \bar{Z}_{i}^{(s)*} \, | \, \mathbf{X} \right ) } \\
  &+ \text{Var} \left (\frac{n_i}{n_i - 1} \bar{Z}_{i}^{(r)*} \bar{Z}_{i}^{(s)*} \, | \, \mathbf{X} \right ) \\
  &\overset{N \rightarrow \infty}{\longrightarrow} 0
  \end{split}
\end{equation}
where the convergence follows from result \eqref{eq:bs-covariance-convergence-partial-result-6} and \eqref{eq:bs-covariance-convergence-partial-result-7}.
Note that this implies \eqref{eq:convergence-variance-bs-covariance} since the factor $N/n_i$ converges to $1/\kappa_i$ for $N \rightarrow \infty$ due to \Atwo{}.
Finally, we apply \eqref{eq:convergence-expectaion-bs-covariance} and \eqref{eq:convergence-variance-bs-covariance} in Chebyshev's inequality
\begin{equation}
  \label{eq:bs-covariance-convergence-chebychev}
  \begin{split}
    {\epsilon^2} P( | \sigma_i^{(rs)*} - \sigma_i^{(rs)} | \geq \epsilon \, | \, \mathbf{X}) %
    &\leq \mathbb{E}((\sigma_i^{(rs)*} - \sigma_i^{(rs)} )^2 \, | \, \mathbf{X}) \\
    &= \mathbb{E}( (\sigma_i^{(rs)*})^2 \, | \, \mathbf{X}) - 2 \sigma_i^{(rs)} \mathbb{E}(\sigma_i^{(rs)*} \, | \, \mathbf{X}) + (\sigma_i^{(rs)})^2 \\
    &= \text{Var}( \sigma_i^{(rs)*} \, | \, \mathbf{X}) + \mathbb{E}( \sigma_i^{(rs)*} \, | \, \mathbf{X})^2 \\
    &- 2 \sigma_i^{(rs)} \mathbb{E}(\sigma_i^{(rs)*} \, | \, \mathbf{X}) + (\sigma_i^{(rs)})^2 \\
    &\overset{P}{\longrightarrow} 0 + (\sigma_i^{(rs)})^2 - 2(\sigma_i^{(rs)})^2 + (\sigma_i^{(rs)})^2 \\
    &= 0
  \end{split}
\end{equation}
which yields the conditional consistency of $\mathbf{S}^*$ in \autoref{theorem:conditional-consistency}.

\subsubsection{Conditional Consistency of the Bootstrap Estimators of Coefficients and of Degrees of Freedom}

We will conduct the consistency proof for $\boldsymbol{\gamma}^*$ and $f^*$ by tracing back to the consistency of the covariance estimator, which we have already shown in the previous subsection.
Note that the consistency of $f^*$ directly follows from the consistency of $\boldsymbol{\gamma}^*$, and thus, we focus on the latter here.

\paragraph{Derivation of the Limit.}

Since Bathke and Brunner~\cite{bathkeNonparametricAlternativeAnalysis2003} did not define the limiting coefficient vector ${\boldsymbol{\gamma} = (\gamma^{(1)}, \ldots, \gamma^{(d)})'}$ explicitly, we derive it here.
Coefficients are obtained as the solution of the system of linear equations that minimizes the variance
$$
{\text{Var}( \sum_{i = 1}^{a} \frac{n_i}{N} (1, -\boldsymbol{\gamma}')\bar{\mathbf{Y}}_i)}
$$
of a weighted sum of adjusted ARTs, where ${\bar{\mathbf{Y}}_i = (\bar{Y}_i^{(0)}, \ldots, \bar{Y}_i^{(d)})'}$.
Recall from \autoref{sec:covariate-adjustment} Bathke and Brunner's consistent estimator $\widehat{\boldsymbol{\gamma}}$~\cite{bathkeNonparametricAlternativeAnalysis2003}:
\DefinitionEstimatorGammaMatrix*

\noindent
Thereby, the entries $\widehat{C}^{(rs)}$ are defined as follows:
\DefinitionEstimatorGammaC*

\noindent
Note that we can rewrite~\eqref{eq:definition-estimator-gamma-c} to obtain an expression consisting of the covariance terms $\widehat{\sigma}_i^{(rs)}$ from~\eqref{eq:definition-empirical-covariance}:
\begin{equation}
  \label{eq:expression-gamma-c-as-sigma}
  \begin{split}
    \widehat{C}^{(rs)} %
    &= \sum_{i = 1}^a \sum_{k = 1}^{n_i} \left ( \frac{n_i}{N} \right )^2 \frac{n_i - 1}{n_i} \frac{N}{n_i(n_i - 1)} \left ( \widehat{Y}_{ik}^{(r)} - \widehat{q}_i^{\, (r)} \right ) \left (\widehat{Y}_{ik}^{(s)} - \widehat{q}_i^{\, (s)} \right ) \\
    &= \sum_{i = 1}^{a} \left ( \frac{n_i}{N} \right )^2 \frac{n_i - 1}{n_i} \widehat{\sigma}_i^{(rs)} 
  \end{split}
\end{equation}
Due to \Atwo{} and the consistency of $\widehat{\sigma}_i^{(rs)}$~\cite{munzelNonparametricMethodsMultivariate2000}, which holds for all $i = 1, \ldots, a$, we have
\begin{equation}
  \label{eq:convergence-empirical-gamma-c}
  \widehat{C}^{(rs)} \overset{P}{\longrightarrow} \sum_{i = 1}^{a} \kappa_i^2 \sigma_i^{(rs)} \eqqcolon C^{(rs)}
\end{equation}
where $\sigma_i^{(rs)}$ is defined in~\eqref{eq:definition-asymptotic-covariance}.
Note that both matrix inversion and matrix multiplication are continuous operations, and thus, due to the continuous mapping theorem, we obtain the following convergence:
\begin{equation}
  \label{eq:definition-gamma-matrix}
  \widehat{\boldsymbol{\gamma}} \overset{P}{\longrightarrow} \begin{pmatrix}
    C^{(11)} & C^{(12)} & \cdots & C^{(1d)} \\
    C^{(21)} & \ddots &  & \vdots \\
    \vdots & & & \\
    C^{(d1)} & \cdots & & C^{(dd)}
  \end{pmatrix}^{-1}
  \begin{pmatrix}
    C^{(01)} \\
    C^{(02)} \\
    \vdots \\
    C^{(0d)}
  \end{pmatrix} \eqqcolon \boldsymbol{\gamma}
\end{equation}
That is, $\boldsymbol{\gamma}$ can be defined in terms of the limiting $C^{(rs)}$; {$r, s = 0, \dots d$}.

\paragraph{Consistency of the Bootstrap Coefficient Estimator.}

Recall the definition of the bootstrap estimator we gave in \autoref{sec:resampling-nancova}:
\DefinitionBsGammaC*

\noindent
As with the non-bootstrap counterpart, we can re-write \eqref{eq:definition-bs-gamma-c} to obtain an expression in terms of the bootstrap covariance estimators $\sigma_i^{(rs)*}$ defined in~\eqref{eq:definition-bs-covariance-estimator}:
\begin{equation}
  \label{eq:expression-bs-gamma-c-as-sigma}
  \begin{split}
    C^{(rs)*} %
    &= \sum_{i = 1}^a \sum_{k = 1}^{n_i} \left ( \frac{n_i}{N} \right )^2 \frac{n_i - 1}{n_i} \frac{N}{n_i(n_i - 1)} \left ( Y_{ik}^{(r)*} - \bar{Y}_i^{(r)*} \right ) \left (Y_{ik}^{(s)*} - \bar{Y}_i^{(s)*} \right ) \\
    &= \sum_{i = 1}^{a} \left ( \frac{n_i}{N} \right )^2 \frac{n_i - 1}{n_i} \sigma_i^{(rs)*} 
  \end{split}
\end{equation}
Thus, we can conclude from~\eqref{eq:bs-covariance-convergence-chebychev}, conditionally on $\mathbf{X}$, it holds that
\begin{equation}
  \label{eq:convergence-bs-gamma-c}
  C^{(rs)*} \overset{P^*}{\longrightarrow} C^{(rs)} \hspace{3mm} \text{in probability}
\end{equation}
Moreover, the continuous mapping theorem implies with respect to our bootstrap estimator $\boldsymbol{\gamma}^*$, conditionally on $\mathbf{X}$, that
\begin{equation}
  \label{eq:convergence-bs-gamma-matrix}
  \begin{split}
    \boldsymbol{\gamma}^* &= \begin{pmatrix}
      C^{(11)*} & C^{(12)*} & \cdots & C^{(1d)*} \\
      C^{(21)*} & \ddots &  & \vdots \\
      \vdots & & & \\
      C^{(d1)*} & \cdots & & C^{(dd)*}
    \end{pmatrix}^{-1}
    \begin{pmatrix}
      C^{(01)*} \\
      C^{(02)*} \\
      \vdots \\
      C^{(0d)*}
    \end{pmatrix} \\
    &\overset{P^*}{\longrightarrow} \boldsymbol{\gamma} \hspace{3mm} \text{in probability}
  \end{split}
\end{equation}

\paragraph{Consistency of the Bootstrap Degrees of Freedom Estimator.}

From \eqref{eq:convergence-bs-gamma-matrix}, we can conclude that also the matrix ${\boldsymbol{\Gamma}^* =  \mathbf{I}_a \otimes ( 1, (-\boldsymbol{\gamma}^*)')}$ converges conditionally on $\mathbf{X}$ as follows:
\begin{equation}
  \label{eq:convergence-bs-auxiliary-gamma-matrix}
  \boldsymbol{\Gamma}^* \overset{P^*}{\longrightarrow} \boldsymbol{\Gamma} \coloneqq \mathbf{I}_a \otimes ( 1, -\boldsymbol{\gamma}') \hspace{3mm} \text{in probability}
\end{equation}
Using the consistency result for $\mathbf{S}^*$ in \autoref{theorem:conditional-consistency}, we also obtain
\begin{equation}
  \label{eq:convergence-bs-adjusted-covariance-matrix}
  \boldsymbol{\Sigma}^* = \boldsymbol{\Gamma}^* \mathbf{S}^* (\boldsymbol{\Gamma}^*)' \overset{P^*}{\longrightarrow} \boldsymbol{\Gamma} \mathbf{S} \boldsymbol{\Gamma}' \eqqcolon \boldsymbol{\Sigma} \hspace{3mm} \text{in probability}
\end{equation}
conditionally on $\mathbf{X}$.
Thus, the consistency of the degrees of freedom estimator $f^*$ directly follows, using the continuity of the involved matrix operations as follows:
\begin{equation}
  \label{eq:convergence-bs-f}
  f^* = \frac%
  {\text{tr}(\mathbf{T} \boldsymbol{\Sigma}^*)^2}%
  {\text{tr}(\mathbf{T} \boldsymbol{\Sigma}^* \mathbf{T} \boldsymbol{\Sigma}^*)} \overset{P^*}{\longrightarrow} \frac%
  {\text{tr}(\mathbf{T} \boldsymbol{\Sigma})^2}%
  {\text{tr}(\mathbf{T} \boldsymbol{\Sigma} \mathbf{T} \boldsymbol{\Sigma})} \eqqcolon f  \hspace{3mm} \text{in probability}
\end{equation}
conditionally on $\mathbf{X}$.

\subsection{Proof for \autoref{theorem:convergence-of-bootstrap-ats}}
\label{subsec:proof-theorem-convergence-of-bootstrap-ats}

In this section, we will provide a proof for the following theorem:

\convergenceOfBootstrapATS*

\noindent
Since we already derived the asymptotic null distribution of $A_N$ in section \autoref{subsec:null-ditribution-an}, it remains to show that the conditional distribution of $A_N^*$ converges to the same limit.
The proof will use both \autoref{theorem:convergence-of-brt} and \autoref{theorem:conditional-consistency}, and moreover, standard results on quadratic forms~\cite{mathaiQuadraticFormsRandom1992}.
Firstly, we recall the definition of $A_N^*$ from \autoref{sec:resampling-nancova}:
\AtsSingleBootstrap*

\noindent
Then, from \autoref{theorem:convergence-of-brt} and \autoref{theorem:conditional-consistency}, we can already conclude, conditionally on $\mathbf{X}$, that
\begin{equation}
  \label{eq:convergence-a-n-star-numerator}
  \sqrt{N} \boldsymbol{\Gamma}^* (\bar{\mathbf{Y}}^* - \widehat{\mathbf{q}}) \overset{d^*}{\longrightarrow} \mathcal{N}(\mathbf{0}, \boldsymbol{\Sigma}) \hspace{3mm} \text{in probability}
\end{equation}
where ${\boldsymbol{\Sigma} = \boldsymbol{\Gamma} \mathbf{S} \boldsymbol{\Gamma}'}$.
Consequently, it also holds, conditionally on $\mathbf{X}$, that
\begin{equation}
  \label{eq:convergence-a-n-star-full}
  A_N^* \overset{d^*}{\longrightarrow} \frac{f}{\text{tr}(\mathbf{T \Sigma})} \sum_{i = 1}^a \lambda_i U_i \hspace{3mm} \text{in probability}
\end{equation}
where $\lambda_1, \ldots, \lambda_a$ are the eigenvalues of $\mathbf{T} \boldsymbol{\Sigma} \mathbf{T}$ and $U_1, \ldots, U_a \overset{\text{iid}}{\sim} \chi_1^2$, which completes the proof.